\documentclass[a4paper,11pt]{article}
\usepackage{jcappub,bm,color} 
\usepackage{amssymb,amsfonts,slashed,amsthm,amsmath,graphicx, soul}
\usepackage[caption=false]{subfig}
\usepackage{comment}
\usepackage{booktabs}
\usepackage{multirow}
\usepackage{makecell}

\newcommand{\newc}{\newcommand}
\newc{\gsim}{\lower.7ex\hbox{$\;\stackrel{\textstyle>}{\sim}\;$}}
\newc{\lsim}{\lower.7ex\hbox{$\;\stackrel{\textstyle<}{\sim}\;$}}
\newc{\gev}{\,{\rm GeV}}
\newc{\mev}{\,{\rm MeV}}
\newc{\ev}{\,{\rm eV}}
\newc{\kev}{\,{\rm keV}}
\newc{\tev}{\,{\rm TeV}}
\newc{\MHT}{$H_T^{\text{miss}}$}
\newc{\MET}{$\slashed{E}_T$}
\newc{\MTT}{$M_{T2}$}

\def\ln{\mathop{\rm ln}}

\newc{\mz}{M_Z}
\newc{\mpl}{M_*}
\newc{\mw}{m_{\rm weak}}
\newc{\nr}[1]{N^c_R{}_{#1}}

\def\beq{\begin{equation}}
\def\eeq{\end{equation}}
\newcommand{\bea}{\begin{eqnarray}\begin{aligned}}
\newcommand{\eea}{\end{aligned}\end{eqnarray}}
\def\bitem{\begin{itemize}}
\def\eitem{\end{itemize}}

\renewcommand{\thefootnote}{\fnsymbol{footnote}}

\definecolor{orange}{RGB}{255,100,0}
\definecolor{rosepink}{RGB}{239,140,159}

\begin{document}

\subheader{{\rm UT-Komaba/23-11}}

\title{Interactions between several types of cosmic strings}

\author[a]{Kohei Fujikura,}
\author[b]{Siyao Li,}
\author[c,b]{and Masahide Yamaguchi}

\affiliation[a]{\it Graduate School of Arts and Sciences, University of Tokyo, Komaba,\\ Meguro-ku, Tokyo 153-8902,
Japan}
\affiliation[b]{\it Department of Physics, Tokyo Institute of Technology,\\Tokyo 152-8551, Japan}
\affiliation[c]{\it Cosmology, Gravity and Astroparticle Physics Group, \\Center for Theoretical Physics of the Universe,\\ Institute for Basic Science, Daejeon 34126, Korea}

\emailAdd{kfujikura@g.ecc.u-tokyo.ac.jp}
\emailAdd{li.s.ap@m.titech.ac.jp}
\emailAdd{gucci@ibs.re.kr}

\abstract{
We study the interaction of several types of static straight cosmic strings, including local strings, global strings, and bosonic superconducting strings with and without magnetic currents.
First, we evaluate the interaction energy of two widely separated cosmic strings using the point source formalism and show that the most dominant contribution to the interaction energy comes from the excitation of the lightest mediator particles in a underlying theory.
The interaction energy at arbitrary separation distances is then analyzed numerically by the gradient flow method.
It turns out that an additional scalar field introduced in the bosonic superconducting string becomes an additional source of attraction.
For such a bosonic superconducting string, we find that a string with two winding numbers is energetically favorable compared to two strings with a single winding number in a certain parameter region.
Our analysis reveals that a phase structure of bosonic superconducting strings is richer than that of local and global strings and that the formation of bound states at intersections of bosonic superconducting strings is favored.
}

\maketitle


\renewcommand{\thefootnote}{\arabic{footnote}}
\setcounter{footnote}{0}


\section{Introduction}

In the early universe, cosmological thermal phase transitions associated with spontaneous symmetry breaking can take place.
If the topology of the vacuum manifold is not simply connected, then after phase transitions, macroscopic one-dimensional line-like topological defects called cosmic strings can be produced through the Kibble-Zurek mechanism~\cite{Kibble:1976sj,Zurek:1985qw}. (See Ref.~\cite{Vilenkin:2000jqa} for the review of cosmic strings.)
A typical example of a (local) cosmic string solution is the Abrikosov-Nielsen-Olesen (ANO) string~\cite{Abrikosov:1956sx,Nielsen:1973cs} realized in the Abelian-Higgs model, where the condensation of the Higgs field takes place.
Further introduction of a new $\widetilde{U}(1)$ gauge group and a new Higgs field leads to a localized condensate of the new Higgs inside the string in a certain parameter region, showing persistent current-carrying superconductivity~\cite{Witten:1984eb}.
In addition, if the theory includes Yukawa coupling between the string-forming Higgs and new fermions, there also exists a current-carrying cosmic string solution~\cite{Jackiw:1981ee,Witten:1984eb}. 
Depending on the spin statistics of the charge carriers, such current-carrying string solutions are conventionally called bosonic or fermionic superconducting strings, and their dynamics and phenomena have been investigated in many literatures~\cite{Lazarides:1984zq,Ostriker:1986xc,Hill:1987qx,Babul:1987me,Amsterdamski:1988zp,Davis:1988jp,Davis:1988jq,Ganoulis:1989hz,Iwazaki:1997bk,Fukuda:2020kym,Agrawal:2020euj,Abe:2020ure,Abe:2022rrh}.

After the formation of cosmic strings, they constitute a web-like structure called a string network system.
During the evolution of a string network system, cosmic strings can intersect many times, and the dynamics of these intersections can affect the fate of the string network system.
In particular, when two cosmic string segments intersect, it is conceivable that they generically snap and reconnect with their partners.
This process is called reconnection (or intercommutation), and occurs dominantly for simple string solutions such as local strings and global strings, unless the relative velocities are sufficiently fast~\cite{Shellard:1987bv,Polchinski:1988cn,Shellard:1988zx,Matzner:1988qqj}.
Interestingly, if the interaction between two string segments is attractive and strong enough, Y-shaped junctions (Y-junctions) can form after string intersection~\cite{Bettencourt:1996qe}. The formation of Y-junctions is discussed in Refs.~\cite{Copeland:2006eh,Copeland:2006if,Salmi:2007ah,Bevis:2008hg,Bevis:2009az,Hiramatsu:2013yxa,Elghozi:2014kya}. Characteristic phenomena of Y-junctions such as distinctive gravitational lensing~\cite{Shlaer:2005ry,Brandenberger:2007ae}, gravitational wave bursts due to cusps and kinks~\cite{Binetruy:2009vt,Binetruy:2010cc,Matsui:2020hzi} have also been discussed.

As a first step toward clarifying the dynamics of string network systems, including the formation of Y-junctions, it is very important to reveal the interaction between two static, straight cosmic string segments.
(Here, ``static" means the velocity of the string is zero.)
Analytical studies of the interaction between two static, parallel global and local strings are initiated in Refs~\cite{Shellard:1987bv,Bettencourt:1994kf} under the assumption that the whole field configurations can be approximated by the superposition of each string.
After these works, Speight proposed a novel method, called point source formalism, to calculate the interaction energy of two local strings~\cite{Speight:1996px}.
It is shown that the asymptotic field configurations of a local string can be well represented by line-like static sources that couple linearly to the fields in the underlying theory.
When two strings are sufficiently separated, the total external sources of the two-string system are given by the superposition of each string's source.
The interaction energy can be then evaluated in linearized field theories using the standard Green's function method.
This formulation can be easily extended to the calculation of interaction energies for more complicated cosmic string solutions.

On the other hand, when two static cosmic string segments are sufficiently close together, the point source formalism is not suitable to estimate the interaction energy.
Clearly, in such a case, the non-linear effects caused by internal structure of cosmic strings on interaction energy cannot be ignored.
Therefore, when two strings are very close together, it is very challenging to compute interaction energy analytically, which results in the need for numerical calculations that do not rely on the point source formalism.
The interaction energy of two local strings at arbitrary separation distances have been accurately evaluated using the variational approach in Ref.~\cite{Jacobs:1978ch} and the gradient flow method in Ref.~\cite{Eto:2022hyt}.
(See also Refs.~\cite{2011PhRvB..83e4516C,2022arXiv220407260S} for related works.)

While the interaction energy and the dynamics of intersections of a simple two-local-string system have been clarified, cosmic string solutions with more complicated field configurations, such as bosonic superconducting strings, have not been fully understood yet.
Ref.~\cite{MacKenzie:2003jp} has studied the interaction energy of bosonic superconducting strings without current, using the point source formalism that takes into account the effect of another scalar field condensing inside the string.
However, there have been neither analytical studies of the interaction energy of bosonic current-carrying strings using the point source formalism nor numerical studies of bosonic superconducting strings that include the effects of nonlinear internal structure.\footnote{Numerical simulations based on field theory for a perpendicular intersection of two current-carrying bosonic superconducting strings have been done in Ref.~\cite{Laguna:1990it}, and the intersection described by the conservative elastic string model, characteristic of current-carrying strings, has been studied in Refs.~\cite{Steer:2017xgh,Rybak:2018oks,Rybak:2020pma}}
One of the main purposes of the present paper is to determine the interaction energy of two static, parallel bosonic superconducting strings using analytic calculations with the point source formalism and numerical calculations with the gradient flow method. In addition, for completeness, we will also apply this to local(ANO) and global strings.

For bosonic superconducting strings, we consider both the cases with and without magnetic currents associated with newly introduced $\tilde{U}(1)$ gauge group.
For the case without magnetic currents, we show that the interaction sourced by the Higgs field in the fundamental representation of $\widetilde{U}(1)$ is always attractive, which gives a dominant contribution to the interaction energy in a certain parameter space.
Especially, the short-range attraction due to the present of this Higgs field results in a non-trivial phase structure of the interaction energy of the bosonic superconducting strings.
We argue that this short-range attraction cannot be seen in the point source formalism and can only be captured by a full numerical calculation.
In the case with magnetic currents, we find that the contribution triggered by the $\widetilde{U}(1)$ gauge field is related to the directions of the currents. 
But its effect is only important in the long-distance limit due to the effect of the back-reaction on $\widetilde{U}(1)$ Higgs field (current quenching effect), which allows only small values of currents.
Meanwhile, the short-range attraction is found to be weaker with the presence of magnetic currents as another result of current quenching effect.
Consequently, this attraction is maximal when the magnetic current is zero.

This paper is organized as follows.
In Sec.~\ref{sec:cosmic string solution}, we review several static cosmic string solutions, including local(ANO) strings, global strings, bosonic superconducting strings.
We then show that asymptotic configurations can be obtained in the linearized field theory by introducing certain external sources.
In Sec.~\ref{sec:interaction}, the interaction energies of various string segments are estimated using the point source formalism.
In Sec.~\ref{sec:numerical study}, we present our numerical results using the relaxation method (gradient flow method).
Sec.~\ref{sec:conclusion} is devoted to conclusions and discussion.
Some numerical details are given in Appendices.

\section{Cosmic string solutions}\label{sec:cosmic string solution}

In this section, we review various types of cosmic string solutions, including local, global, bosonic superconducting cosmic strings.
We derive the asymptotic configurations of these strings analytically in the long-distance limit.
It is shown that the asymptotic field configurations can be obtained by introducing corresponding external sources that linearly couple to fields in the free field theory.

\subsection{Local strings}\label{sec:local strings}

Let us first consider the simplest cosmic string solution realized in the Abelian-Higgs model, known as the Abrikosov-Nielsen-Olsen (ANO) string~\cite{Abrikosov:1956sx,Nielsen:1973cs}.
The Lagrangian density of the Abelian-Higgs model is given by\footnote{Here, we focus on the conventional polynomial form of the potential. There exist string solutions with the Coleman-Weinberg type potentials~\cite{Coleman:1973jx}. Its string solution and interaction energy between these strings have recently been investigated in details by Ref.~\cite{Eto:2022hyt}.}
\begin{equation}
\begin{split}
&\mathcal{L}_{\rm AH} = -\frac{1}{4}F^{\mu\nu}F_{\mu\nu} + |D_\mu \phi|^2 - V(\phi),\\
&V(\phi) = \frac{\lambda_\phi}{4}\left(|\phi|^2-\eta_\phi^2\right)^2. \label{eq:Abelian-Higgs}
\end{split}
\end{equation}
where $F_{\mu\nu}=\partial_\mu A_\nu -\partial_\nu A_\mu$ and $D_\mu \equiv \partial_\mu - ie A_\mu$ are the field strength and the covariant derivative of the $U(1)$ gauge field, respectively.
The variational principle leads to the following equations of motion:
\begin{equation}
\begin{split}
&\left(D_\mu D^\mu- \frac{\lambda_\phi}{2}\left(|\phi|^2 -\eta_\phi^2\right)\right) \phi =0, \\
&\partial_\nu F^{\nu\mu} = J^\mu .\label{eq:ANO EOM}
\end{split}
\end{equation}
Here, $J^\mu$ is the Noether current associated with local $U(1)$ symmetry given by
\begin{align}
J_\mu =i e \left(\phi D_\mu \phi^* - \phi^* D_\mu \phi \right).\label{eq:noether current}
\end{align}

The ANO string solution is obtained by considering the boundary conditions with non-trivial windings.
Assuming that the cosmic string configuration is static, straight (here and hereafter we use ``straight" to indicate translational symmetric in the string direction) and circularly symmetric on the plane perpendicular to the string axis direction, we can use the following ansatz:
\begin{align}
\phi(x) = \phi_r(r)e^{in \theta},~~  A_\mu(x) = A_\theta (r) \delta^\mu_{\,\theta} . \label{eq:ansatz}
\end{align}
Here, $(r,~\theta,~z)$ is cylindrical coordinate, where $r,~\theta$ and $z$ is the radial distance, the angle and the height, respectively.
$\phi_r(r)$ and $A_\theta(r)$ are real functions, and $n$ is a non-zero integer number called winding number.
Boundary conditions of $\phi$ and $A_\mu$ at infinity are determined by finiteness of the energy, that is, the scalar potential, $V(\phi)$, and the covariant derivative $D_\mu \phi$ must vanish at infinity:
\begin{align}
\phi_r =\eta,~~D_\mu \phi =0,~~r\to \infty . \label{eq:boundary condition}
\end{align}
Regularities of $|\phi|$ and $A_\theta$ at the origin require the following boundary conditions,
\begin{align}
	\phi_r(r\to 0)= 0,~A_\theta (r\to 0) = 0. \label{eq:ANO boundary conditions at the origin}
\end{align}
Analytic solutions of Eq.~\eqref{eq:ANO EOM} under boundary conditions Eqs~\eqref{eq:boundary condition} and \eqref{eq:ANO boundary conditions at the origin} are not known, and hence, a numerical calculation is required.
See appendix~\ref{appendix:numerical calculation} for detailed setup of numerical calculations.
A consequence of boundary condition Eq.~\eqref{eq:boundary condition} is that a magnetic flux must be quantized at infinity.
\begin{align}
	 \oint_{{\mathbb S}^1} A  = \dfrac{n}{e}\oint_{{\mathbb S}^1}  \mathrm{d}\theta =\dfrac{2\pi n}{e},~(r\to \infty). \label{eq:magnetic flux}
\end{align}
Here, $A\equiv A_\mu \mathrm{d}x^\mu$ while the integration is taken over the circle enclosing the origin in the two-dimensional plane which is perpendicular to the straight ANO string.
Note that the first equality only holds at infinity.

Let us now focus on the asymptotic solutions.
We work in the different field parameterizations defined by
\begin{align}
\phi (x) = \left(\eta +\frac{\sigma(x)}{\sqrt{2}} \right) e^{i \pi(x)},~U_\mu (x)=A_\mu (x) -\frac{1}{e} \partial_{\mu} \pi(x), \label{eq:field parametrizations}
\end{align}
where $\sigma (x)$ is the radial component of the $\phi$ field and $\pi(x)$ is the Nambu-Goldstone boson (NG-boson).
Although the above field parameterizations are ill-defined at the origin $\phi = 0$, they are convenient to seek asymptotic field configurations of the ANO string at infinity, as we will see below.
With these field parameterizations, the Lagrangian density Eq.~\eqref{eq:Abelian-Higgs} can be rewritten as 
\begin{align}
\mathcal{L}_{\rm AH} &= \frac{1}{2} \partial_\mu \sigma \partial^\mu \sigma -\frac{1}{2}m_{\phi}^2 \sigma^2 -\frac{1}{4}\mathcal{F}^{\mu\nu}\mathcal{F}_{\mu\nu} +\frac{1}{2} m_e^2 U_\mu U^\mu \nonumber \\
&+\sqrt{2}e^2 \eta \sigma U_\mu U^\mu +\frac{e^2}{2}\sigma^2 U_\mu U^\mu - \frac{\sqrt{2}\lambda}{4} \eta \sigma^3-\frac{\lambda}{16}\sigma^4, \label{eq:unitary gauge L}
\end{align}
where $m_\phi \equiv \sqrt{\lambda_\phi}\eta$ and $m_e\equiv \sqrt{2}e \eta$ are masses of the scalar and $U(1)$ gauge field evaluated at infinity, respectively, and $\mathcal{F}_{\mu\nu} = \partial_\mu U_\nu -\partial_\nu U_\mu = F_{\mu\nu}$.

Let us find asymptotic solutions for the scalar and the gauge fields.
Circular symmetry restricts forms of field configurations as $\sigma (x)= \sigma (r)$ and $U_\mu(x) =U_\theta(r)\delta^\mu_{\,\theta}$.
Since $\sigma (r) \simeq 0$ due to the boundary conditions Eq.~\eqref{eq:boundary condition}, we can neglect terms that depend on $\sigma(r)$ at infinity.
Then the equation of motion for $U_\theta(r)$ becomes linear and is given by
\begin{align}
\left( \frac{\partial^2}{\partial r^2}-\frac{1}{r}\frac{\partial }{\partial r}-m_e^2\right)U_\theta(r)=0 \label{eq:eom1}.
\end{align}
Therefore, we can obtain an analytical asymptotic solution of $U_\theta(r)$ under boundary condition Eq.~\eqref{eq:boundary condition} as 
\begin{align}
U^{\rm sol}_\theta(r) = k_er K_1 (m_e r) ,\label{eq:asymptotic U}
\end{align}
where $K_m (x)$ is the modified Bessel function of the second kind order $m$, while $k_e$ is a numerical constant.
It is worth noting that the modified Bessel functions have the following properties,
\begin{align}
	K_0(r) = \sqrt{\dfrac{\pi}{2r}}e^{-r},~K_1(r) = \sqrt{\dfrac{\pi}{2r}}e^{-r},~(r\to\infty).
\end{align}
Since $K_1(x)$ is exponentially suppressed at large distance $x\gg 1$, we can safely neglect non-linear terms of $U_\theta$, and then, the equation of motion for $\sigma (r)$ is simplified as 
\begin{align}
\left(\frac{\partial^2}{\partial r^2} +\frac{1}{r} \frac{\partial}{\partial r} -m_\phi^2\right)\sigma(r)=0 \label{eq:eom2}.
\end{align}
Here, we have assumed that the quadratic term of the gauge field $(U_\theta)^2$ is negligible compared to $\sigma (r)$, which is justified only for $\lambda < 8e^2$~\cite{Perivolaropoulos:1993uj}.
The solution consistent with the boundary condition Eq.~\eqref{eq:boundary condition} is given by
\begin{align}
\sigma^{\rm sol}(r) = k_\phi K_0 (m_\phi r), \label{eq:asymptotic sigma}
\end{align}
where $k_\phi$ is a constant.
For the parameter region $\lambda > 8e^2$, the quadratic term $(U_\theta)^2$ is not negligible, and thus, the asymptotic expression Eq.~\eqref{eq:asymptotic sigma} is no longer justified~\cite{Perivolaropoulos:1993uj}.
In this case, the $(U_\theta)^2$ term has to be included to derive the correct asymptotic configurations.

Numerical constants $k_\phi$ and $k_e$ are determined by the requirement that exact numerical solutions of Eq.~\eqref{eq:ANO EOM} in the long-distance limit should coincide with the asymptotic solutions.
For the ANO string configuration, this procedure is explicitly examined in Ref.~\cite{Speight:1996px}. 
For a special case $\lambda/ 2e^2=1$ called the critical coupling or Bogomol’nyi-Prasad-Sommerfield (BPS) limit, analytic expressions have been found~\cite{Bettencourt:1994kf}.

Let us next find external sources that represent the presence of an ANO string with given constants $k_\phi$ and $k_e$.
Following Ref.~\cite{Speight:1996px}, we introduce external sources $J_\sigma$ and $j_\mu$ in the linearized field theory as 
\begin{align}
\mathcal{L}_{\rm AH} = \frac{1}{2}\partial_\mu \sigma \partial^\mu \sigma -\frac{1}{2}m_\phi^2 \sigma^2 -\frac{1}{4}\mathcal{F}^{\mu\nu}\mathcal{F}_{\mu\nu} +\frac{1}{2}m_e^2 U_\mu U^\mu - J_\sigma \sigma -j_\mu U^\mu.\label{eq:total L1}
\end{align}
At infinity, since string configurations are described by Eqs.~\eqref{eq:asymptotic U} and \eqref{eq:asymptotic sigma}, we can read off the explicit forms of external sources:
\begin{align}
&J_\sigma \equiv \left( \nabla^2-m_\phi^2\right) \sigma^{\rm sol} (|{\bm x}|)= -2\pi k_\phi\delta^{(2)} ({\bm x}),\label{eq:scalar source}\\
&j_i \equiv \left(\nabla^2 -m_e^2  \right) U^{\rm sol}_i (|{\bm x}|) =  2\pi \frac{k_e}{m_e}  \epsilon_{ji}\nabla_j  \delta^{(2)}({\bm x})\label{eq:magnetic source}.
\end{align}
Here, we change the two-dimensional coordinate from $(r,\theta)$ to the two-dimensional Cartesian coordinate ${\bm x}\equiv (x,y)$ for later convenience.
The Laplace operator is defined by $\nabla^2\equiv \partial_{x}^2+\partial_{y}^2$, while $\epsilon_{ij}$ is the two-dimensional Levi-Civita symbol.
$\delta^{(2)}({\bm x})$ is the two-dimensional delta function.
Following relations are used to derive Eq.~\eqref{eq:magnetic source}
\begin{align}
(\nabla^2-1)K_0 (|{\bm x}|) = - 2\pi\delta^{(2)}({\bm x}),~
\frac{\mathrm{d}}{\mathrm{d}r} K_0 (r) = -K_1 (r).
\end{align}
The above result implies that the ANO string configurations $\sigma (r)$ and $U_\theta (r)$ can be successfully represented by point-like external sources in the long-distance limit where the internal structure of the string is coarse-grained.
As argued in Ref.~\cite{Speight:1996px}, $J_\sigma$ and $j_{\mu}$ represent a scalar monopole with charge $-2\pi k_\phi$ and a magnetic dipole with moment $-2\pi k_e/m_e$ whose direction is parallel or antiparallel to the $z$-direction depending on the sign of the winding number, respectively.
In particular, the existence of magnetic dipole moments follows from the presence of the magnetic flux inside the string given by Eq.~\eqref{eq:magnetic flux}.
Analogous to a charged electron linearly coupled to a massless photon, $J_\sigma$ and $j_{\mu}$ couple to a massive scalar $\sigma$ and a gauge field $A_{\mu}$, respectively, for a local string. This picture is very useful when calculating the interaction between two strings, as seen in Sec.~\ref{sec:interaction}.

\subsection{Global strings}\label{section:global string solution}

In this subsection, we derive solutions for global strings and show that the existence of global strings can be approximately described by external currents associated with scalar and topological charges. The scalar and topological charges are linearly coupled to a massive scalar field and a two-index antisymmetric tensor field, respectively.

The Lagrangian density of the Goldstone model with a complex scalar field is described by
\begin{align}
\mathcal{L}_{\rm global} = |\partial_\mu \phi|^2- V(\phi),
\end{align}
where $V(\phi)$ is defined as same as Eq.~\eqref{eq:Abelian-Higgs}.
With the scalar field parametrization given by Eq.~\eqref{eq:field parametrizations}, above Lagrangian can be rewritten as
\begin{align}
\mathcal{L}_{\rm global} = \frac{1}{2} \partial_\mu \sigma \partial^\mu \sigma  +|\phi|^2 \partial_\mu \pi \partial^\mu \pi -\frac{1}{2} m^2_\phi\sigma^2 -\frac{\sqrt{2}}{4}\lambda_\phi \eta_\phi \sigma^3 -\frac{\lambda_\phi}{16}\sigma^4.
\end{align}
By using the ansatz of the static, straight string solution given by Eq.~\eqref{eq:ansatz}, we obtain the equations of motion of $\sigma$ and $\pi$ fields as
\begin{align}
\left(\frac{\partial^2}{\partial r^2} +\frac{1}{r}\frac{\partial}{\partial r}-m_\phi^2\right) \sigma &-\frac{3}{4}\sqrt{2}\lambda_\phi \eta_\phi \sigma^2 -\frac{\lambda_\phi}{4}\sigma^3 +\sigma(\partial_\mu\pi)^2 +\sqrt{2}\eta_\phi(\partial_\mu \pi)^2 =0,
\label{eq:massive eom}\\
&\partial_\mu(|\phi|^2\partial^\mu\pi)=0.\label{eq:massless eom}
\end{align}
Assuming the power law dependence, $\sigma (r)\propto r^{-a}~(a>0)$, at infinity and inserting $\pi =n\theta$ into above expression, we obtain an asymptotic solution of $\sigma(r)$ at the leading order,
\begin{align}
\sigma^{\rm sol} (r) = -\frac{\sqrt{2}n^2\eta_\phi}{m_\phi^2 r^2} .\label{eq:global config}
\end{align}
We have confirmed that this result is in agreement with the numerical one.

We first focus on the external source describing the massive excitation parameterized by $\sigma(r)$.
We introduce the external source as 
\begin{align}
\mathcal{L} = \frac{1}{2}(\partial_\mu \sigma )^2 -\frac{1}{2}m_\phi^2 \sigma^2 - J_G\sigma.
\end{align}
$J_G$ is also determined by the requirement that it reproduces the scalar configuration given by Eq.~\eqref{eq:global config}.
Substituting the asymptotic solution in Eq.~\eqref{eq:global config}, to the leading order, one obtains
\begin{align}
J_G =  \frac{\sqrt{2}n^2\eta_\phi}{r^2}.
\end{align}
From the above expression, it is obvious that $J_G$ is not a localized source, which is different from the point-like $J_\sigma$ in the ANO string.
This stems from the fact that a contribution from the NG-boson, which is a massless degree of freedom originated from the term $(\partial_\mu \pi)^2$ in Eq.~\eqref{eq:massive eom}, is dominant at infinity.
For the local string, this contribution is canceled by the gauge field configuration.

We now turn our attention to external sources describing the massless excitations of the global string.
Since a non-trivial winding at the center of the string implies the existence of a magnetic flux associated with a NG-boson, it is appropriate to introduce a dual field of NG bosons~\cite{Kalb:1974yc,Vilenkin:1986ku,Davis:1988rw}.
The dual field of the massless real scalar field $\pi(x)$ is a two-index antisymmetric tensor field $B_{\mu\nu}$ satisfying the following equation:
\begin{align}
|\phi|^2\partial_\mu \pi = \frac{\eta_\phi}{2} \epsilon_{\mu\nu\rho\sigma} \partial^\nu B^{\rho\sigma}.\label{eq:on-shell}
\end{align}
This equation makes sense when $|\phi| \ne 0$ and $\pi(x)$ is equivalent to $B_{\mu\nu}$ in Minkowski spacetime as long as the on-shell condition Eq.~\eqref{eq:on-shell} is satisfied~\cite{Kalb:1974yc}.
Then the Lagrangian density in terms of $B_{\mu\nu}$ can be written as
\begin{align}
\mathcal{L}_{\rm dual} &= \frac{\eta_\phi^2}{6|\phi|^2}(H_{\mu\nu\rho})^2, \label{eq:KR action}\\
H^{\mu\nu\rho} &\equiv \partial^\mu B^{\nu\rho}+ \partial^\rho B^{\mu\nu} +\partial^\nu B^{\rho \mu},
\end{align}
where we have omitted the terms only depending on $\sigma$ because they do not play an important role in the following discussion.
$H^{\mu\nu\rho}$ is the field strength tensor of $B_{\mu\nu}$.

In the dual picture, the interaction of NG-boson and its magnetic charge can be equivalently expressed by introducing the electric source of the dual field,
\begin{align}
\mathcal{L} = \mathcal{L}_{\rm dual} - B_{\mu\nu}J^{\mu\nu}_{B}. \label{eq:topological coupling}
\end{align}
Here, $J_B^{\mu\nu}$ is the external source of $B_{\mu\nu}$.
The equation of motion of the dual field is given by 
\begin{align}
\partial_\rho\left(\frac{\eta_\phi^2}{|\phi|^2} H^{\rho\mu\nu} \right)= -J^{\mu\nu}_B. \label{eq:eom dual B}
\end{align}
We can also obtain the equation of motion of $B_{\mu\nu}$ by simply substituting the on-shell condition in Eq.~\eqref{eq:on-shell} into field equation of $\pi$ in Eq.~\eqref{eq:massless eom}.
Then by comparing with Eq.~\eqref{eq:eom dual B} one can read off the expression for $J_{B}^{\mu\nu}$ as
\begin{align}
J^{\mu\nu}_B=\eta_\phi \epsilon^{\mu\nu\rho\alpha}\partial_\rho \partial_\alpha \pi.
\end{align}
Due to the antisymmetric property of $\epsilon^{\mu\nu\rho\sigma}$, the right-hand side of the above equation is zero, except at the origin, $|\phi| = 0$.
At the origin, $\pi$ is not well-defined, therefore, one cannot conclude that $J^{\mu\nu}_B$ vanishes at the origin.
The presence of the magnetic charge of the NG-boson can be seen by integrating over the two-dimensional plane, which is perpendicular to the straight string axis, as 
\begin{align}
    \int \mathrm{d}^2 x\epsilon^{03ij}\partial_i\partial_j\pi=\oint_{{\mathbb S}^1}\mathrm{d}\pi = 2\pi n. \label{eq:NG source integral}
\end{align} 
We see that $J_{B}^{\mu\nu}$ is zero except at the origin, and integrating over a two-dimensional plane gives a nonzero finite value. This is exactly the definition of a delta function, and consequently we have
\begin{align}
J^{03}_{B}({\bm {x}}) = -J^{30}_B({\bm x}) = 2\pi n\eta_\phi\delta^{(2)}({\bm x}). \label{eq:topological source}
\end{align}
Since this external source represents the winding number of the global string, the interaction Eq.~\eqref{eq:topological coupling} with the external source Eq.~\eqref{eq:topological source} can be understood as a topological coupling between the NG-boson and the global string.

\subsection{Bosonic Superconducting strings}\label{sec:bosonic superconducting}

In this subsection, we review the bosonic superconducting string and derive its asymptotic solutions.
The simplest model that accommodates bosonic superconducting solutions possesses two gauge symmetries with $U(1)\times \widetilde{U}(1)$ and the Lagrangian density is defined by~\cite{Witten:1984eb}
\begin{equation}
\begin{split}
    &\mathcal{L} = -\frac{1}{4}F^{\mu\nu}F_{\mu\nu} + |D_\mu \phi|^2 -\frac{1}{4}\Tilde{F}^{\mu\nu}\Tilde{F}_{\mu\nu} +|\Tilde{D}_\mu \Tilde{\phi}|^2 - V(\phi,\Tilde{\phi}),\\
    &V(\phi,\Tilde{\phi}) = \frac{\lambda_\phi}{4} \left( |\phi|^2-\eta_\phi^2 \right)^2 + \frac{\lambda_{\Tilde{\phi}}}{4}\left(|\Tilde{\phi}|^2-\eta_{\Tilde{\phi}}^2\right)^2+\beta |\phi|^2 |\Tilde{\phi}|^2 , \label{eq:superconducting lagrangian density}
\end{split}
\end{equation}
where $F_{\mu\nu}=\partial_\mu A_\nu -\partial_\nu A_\mu $, $\Tilde{F}_{\mu\nu}=\partial_\mu \Tilde{A}_\nu -\partial_\nu \Tilde{A}_\mu$ and $D_\mu \equiv \partial_\mu - ie A_\mu $, $\Tilde{D}_\mu \equiv \partial_\mu - ig\Tilde{A}_\mu$ are the field strength and the covariant derivative of the $U(1)$, and $\widetilde{U}(1)$ gauge field, respectively.
Here, we simply assume $\lambda_\phi > 0,~\lambda_{\Tilde{\phi}}>0$ and $\beta >0$.
The bosonic superconducting solution is realized when $U(1)$ symmetry is spontaneously broken and gives rise to the ANO string configurations, as we discussed in Sec.~\ref{sec:local strings}, while $\widetilde{U}(1)$ symmetry is only broken in the string interior by the localized $\Tilde{\phi}$ condensation.
Thus $|\phi|= 0$ and $|\Tilde\phi| \simeq \eta_{\Tilde{\phi}}$ at the string center, and $|\phi|=\eta_\phi$ and $|\Tilde\phi|=0$ at infinity are realized.

We shall discuss the parameter region leading to the bosonic superconducting cosmic string solution.
For an illustrative purpose, we here neglect the $\Tilde{U}(1)$ gauge field and the non-trivial $z$-dependence of the phase of $\Tilde{\phi}$, which will be included later.
A presence of the biquadratic interaction of the form $\beta |\phi|^2|\Tilde{\phi}|^2$ contributes to the effective mass term of $\Tilde{\phi}$ as
\begin{align}
	m^2_{\Tilde{\phi}}=\beta|\phi|^2-\dfrac{\lambda_{\Tilde{\phi}}}{2}\eta^2_{\Tilde{\phi}}+\lambda_{\Tilde{\phi}}|\tilde{\phi}|^2. \label{eq:effective mass term}
\end{align}
$\widetilde{U}(1)$ symmetry is unbroken well outside the string when the square-mass of $\Tilde{\phi}$ is positive, that is, $\beta \eta_\phi^2-\lambda_{\Tilde{\phi}} \eta_{\Tilde{\phi}}^2/2>0$.
Also, $\lambda_\phi\eta_\phi^4 > \lambda_{\Tilde{\phi}} \eta_{\Tilde{\phi}}^4$ is required to avoid the metastability at the minimum of potential energy, $|\phi|=\eta_\phi$ and $|{\Tilde{\phi}}|=0$, corresponding to the string exterior.
Near the center of the string, $|\phi| \simeq 0$, condensation $|{\Tilde{\phi}}|\neq 0$ occur around the center of the string due to the instability $m_{\Tilde{\phi}}^2<0$ at $\Tilde{\phi}=0$.
In this argument, however, we disregard the contribution from the gradient energy of $\Tilde{\phi}$.
It is shown in Refs.~\cite{Witten:1984eb,Haws:1988ax,Lemperiere:2002en} that the localized condensation of $\Tilde{\phi}$ is energetically favorable for $\beta \lesssim\lambda_{\Tilde{\phi}}^2\eta_{\tilde{\phi}}^4/(4\lambda_\phi \eta_\phi^4)$ including the gradient energy under the approximation that $\phi(r)\simeq r^n$ around $r=0$ and the back-reaction of $\Tilde{\phi}$ on $\phi$ is negligible.
As a result, conditions to form the bosonic superconducting string without effects of $\tilde{U}(1)$ gauge field are summarized as
\begin{align}
\frac{\lambda_{\Tilde{\phi}}}{2}\dfrac{\eta_{\Tilde{\phi}}^2}{\eta_\phi^2} <\beta \lesssim\frac{1}{4}\frac{\lambda^2_{\Tilde{\phi}}}{\lambda_\phi}\dfrac{\eta_{\Tilde{\phi}}^4}{\eta_\phi^4}. \label{eq:parameter space}
\end{align}

In this class of models, the cosmic string can carry a current flowing along the string and exhibit superconductivity when the phase of $\Tilde{\phi}$ has a non-trivial $z$-dependence.
For a static straight string, the general ansatz of ${\Tilde{\phi}}$ and $\Tilde{A}_\mu$ are given by~\cite{Alford:1990ur}
\begin{align}
    \Tilde{\phi} = \Tilde{\phi}_r(r) e^{-is(r)\alpha(z)},~
    \Tilde{A_\mu} =  -\frac{1}{g}\alpha(z)\partial_\mu s(r), \label{eq:collective coordinates}
\end{align}
where $s$ and $\alpha$ are real functions.
This parameterization is not a mere gauge transformation from $\Tilde{\phi}=\Tilde{\phi}_r$ and $\Tilde{A}_\mu =0$ unless $\alpha(z)$ is independent of $z$.
Therefore it describes the real physical excitation.
This ansatz is equivalent to the following expression up to the gauge transformation,
\begin{align}
    {\Tilde{\phi}} = {\Tilde{\phi}}_r(r),~\Tilde{A}_\mu = \frac{1}{g}s(r)\partial_\mu \alpha(z) . \label{eq:modulate gauge transformation}
\end{align}

With Eq.~\eqref{eq:modulate gauge transformation}, one obtains the equation of motion of $\alpha(z)$ from the Lagrangian density Eq.~\eqref{eq:superconducting lagrangian density},
\begin{align}
       r\,\partial_r s(r)\,\partial_z^2\alpha(z)=0\,.
\end{align}
When $s(r)$ is not independent of $r$, the static solution of this equation is simply given by $\alpha(z)=\omega z$ where $\omega$ is a constant.
With $\phi$ and $A_\mu$ following the same ansatz as the ANO string in Eq.~\eqref{eq:ansatz}, the equations of motion of $\phi_r,~A_\theta,{\Tilde{\phi}}_r$ and $s$ are given by
\begin{equation}
\begin{split}
    &\frac{\partial^2 \phi_r}{\partial r^2} + \frac{1}{r}\frac{\partial \phi_r}{\partial r} - \frac{e^2}{r^2}\phi_r\left( A_\theta - \frac{n}{e} \right)^2 - \frac{1}{2}\lambda_\phi \phi_r (\phi_r^2 - \eta_\phi^2) - \beta|{\Tilde{\phi}}|^2\phi_r = 0,\\
    &\frac{\partial^2 A_\theta}{\partial r^2} - \frac{1}{r}\frac{\partial A_\theta}{\partial r} - 2e^2 \phi_r^2 \left( A_\theta - \frac{n}{e} \right) = 0,\\
    &\frac{\partial^2 \Tilde{\phi}_r}{\partial r^2} + \frac{1}{r}\frac{\partial \Tilde{\phi}_r}{\partial r} - \Tilde{s}^2{\Tilde{\phi}}_r - \frac{1}{2} \lambda_{\Tilde{\phi}}{\Tilde{\phi}}_r({\Tilde{\phi}}_r^2 - \eta_{\Tilde{\phi}}^2) - \beta|\phi|^2{\Tilde{\phi}}_r = 0,\\
    & \frac{\partial^2 \Tilde{s}}{\partial r^2} + \frac{1}{r}\frac{\partial \Tilde{s}}{\partial r} - 2g^2\Tilde{s}{\Tilde{\phi}}_r^2 = 0. \label{eq:bosonic eom}
\end{split}
\end{equation}
Here, we use the normalization $\tilde{s}(r)=\omega s(r)$ for convenience.
Boundary conditions at infinity are determined by the finiteness of the energy, and are given by 
\begin{align}
    &\phi_r(r\to \infty) = \eta_\phi, ~A_\theta (r\to \infty)=\frac{n}{e},~{\Tilde{\phi}}_r (r\to \infty) = 0. \label{eq:superconducting string boundary condition}
\end{align}
The first and the second conditions are the same as those of the ANO string in Eq.~\eqref{eq:boundary condition}.
The regularity at the origin requires 
\begin{align}
    &\phi_r(r \to 0) = 0, ~A_\theta (r \to 0)=0,~{\partial_r \Tilde{\phi}}_r (r \to 0) = 0,~\partial_r \tilde{s}(r \to 0)=0.  \label{eq:superconducting string boundary condition at origin}
\end{align}
As in the case of the ANO string, no exact analytic solution is known. The numerical solutions of Eqs.~\eqref{eq:bosonic eom} are shown in App.~\ref{appendix:numerical calculation}.

Let us now seek the asymptotic solution for the current-carrying superconducting string.
Assuming ${\Tilde{\phi}}_r$ falls faster than $\Tilde{s}(r)^2$ at large $r$, we neglect the third term and the non-linear terms of ${\Tilde{\phi}}_r$ in the equation of motion for ${\Tilde{\phi}}$ in Eq.~\eqref{eq:bosonic eom}. Noting that $\Tilde{s}(r \rightarrow \infty)\simeq \Tilde{s}(r=0)$ contributes to the effective mass of ${\Tilde{\phi}}$, but we do not know the value of $\Tilde{s}$ at infinity, Eq.~\eqref{eq:bosonic eom} can be approximated by
\begin{align}
    \frac{\partial^2 \Tilde{\phi}_r}{\partial r^2} + \frac{1}{r}\frac{\partial \Tilde{\phi}_r}{\partial r} - m_{\Tilde{\phi},\infty}^2\, {\Tilde{\phi}}_r = 0, ~~~~m_{\Tilde{\phi},\infty}^2 \equiv \beta\eta_\phi^2 - \frac{1}{2}\lambda_{\Tilde{\phi}} \eta_{\Tilde{\phi}}^2+\Tilde{s}^2(r=0),
\end{align}
and its solution at infinity is given by
\begin{align}
    {\Tilde{\phi}}_r^{\rm sol} = k_{\Tilde{\phi}} K_0(m_{\Tilde{\phi},\infty} r), \label{eq:asymptotic tildephi}
\end{align}
where $k_{\Tilde{\phi}}$ is a numerical constant.
Since ${\Tilde{\phi}}_r$ is exponentially suppressed at long distance, the $\Tilde{s}{\Tilde{\phi}}^2_r$ term in Eq.~\eqref{eq:bosonic eom} can be ignored. As a result, the equation becomes the standard Laplace equation in two dimensions without angular dependence, whose solution is given by
\begin{align}
	\Tilde{s}^{\rm sol}(r) = k_s \ln \left(\frac{r}{\delta} \right) +k_{s0}.\label{eq:asymptotic gauge field}
\end{align}
Here, $k_s$ and $k_{s0}$ are numerical constants.
$\delta$ is the typical width of string core.
We confirm that these asymptotic solutions in Eqs.~\eqref{eq:asymptotic tildephi} and \eqref{eq:asymptotic gauge field} agree with our exact numerical calculations.
Two numerical constants $k_{\Tilde{\phi}}$ and $k_s$ are then determined by the fitting of numerical calculations shown in table~\ref{table:charge fitting}.

We now briefly explain the effect of $\widetilde{U}(1)$ current induced by the condensation of $\tilde{\phi}$.
The total amount of the current associated with $\widetilde{U}(1)$ symmetry flowing along the straight string is defined by
\begin{align}
    \widetilde{J}_{\rm tot}=\int {\mathrm d}^2x\,\widetilde{J}_z,~\widetilde{J}_z(r)\equiv - 2g  \Tilde{s}(r) \Tilde{\phi}_r^2(r).\label{eq:superconducting current}
\end{align}
Here, $\widetilde{J}_z$ is the $z$-component of the Noether current associated with $\widetilde{U}(1)$.
Since $\widetilde{\phi}_r$ condensation is strongly localized around the string, $\widetilde{J}_z (r)$ is trapped inside the string.
Very interestingly, one cannot make $\widetilde{J}_{\rm tot}$ arbitrarily large due to the following reason.
It is clear that the effective mass term of $\Tilde{\phi}$ defined by Eq.~\eqref{eq:effective mass term} is now modified by the non-zero $\Tilde{s}$ as
\begin{align}
	m_{\Tilde{\phi}}^2 (r)=\beta\phi_r^2(r) - \frac{1}{2}\lambda_{\Tilde{\phi}}\eta_{\Tilde{\phi}}^2+\lambda_{\Tilde{\phi}}\tilde{\phi}_r^2(r)+\Tilde{s}^2(r).
\end{align}
For a large value of $\Tilde{s}(r=0)$, the effective mass term at the origin becomes positive even with $\Tilde{\phi}_r = 0$ near the string center, which results in symmetry restoration of $\widetilde{U}(1)$ and no condensation of $\Tilde{\phi}$.
Hence, for a large amplitude of $\Tilde{s}(r=0)$, a back-reaction to $\Tilde{\phi}$ is significant which makes the strength of the condensation weaker.
Consequently, $\widetilde{J}_{\rm tot}$ begins to get smaller as $\Tilde{s}(r=0)$ is sufficiently large to have significant back rection on $\Tilde{\phi}_r$.
This phenomenology is known to be current quenching~\cite{Witten:1984eb}.
On the other hand, $\widetilde{J}_{\rm tot}$ also becomes smaller for a smaller $\Tilde{s}(r=0)$ when it is small enough that the back-reaction from $\Tilde{s}$ on $\Tilde{\phi}_r$ is a negligible amount. 
Therefore, there exists a maximum of the total current flowing inside the string, at least in the case of the magnetic current.

Next, let us find external sources that represent the bosonic superconducting string.
As in the case of the ANO string, we introduce currents $J_{\Tilde{\phi}}$ and $\Tilde{j}_\mu$ that reproduce the asymptotic field configurations ${\Tilde{\phi}}^{\rm sol}_r$ and $\Tilde{A}_\mu^{\rm sol} \left(\Leftrightarrow \Tilde{s}^{\rm sol}\right)$ in the linear theory,
\begin{align}
    &\mathcal{L} = \mathcal{L_{\rm AH}}+\frac{1}{2}(\partial_\mu{\Tilde{\phi}}_r)^2 - \frac{1}{2}m_{\Tilde{\phi},\infty}^2{\Tilde{\phi}}_r^2 - \frac{1}{4}\Tilde{F}^{\mu\nu}\Tilde{F}_{\mu\nu}-J_{\Tilde{\phi}}{\Tilde{\phi}}_r - \Tilde{j}_\mu \Tilde{A}^\mu. \label{eq:sc interaction Lagrangian}
\end{align}
In this expression, $\mathcal{L}_{\rm AH}$ is defined by Eq.~\eqref{eq:total L1}.
Substituting the asymptotic solutions into the equations of motion derived from Eq.~\eqref{eq:sc interaction Lagrangian}, one can read off corresponding  sources as
\begin{equation}
\begin{split}
    &J_{\Tilde{\phi}} = ( \nabla^2- m_{\Tilde{\phi},\infty}^2){\Tilde{\phi}}_r^{\rm sol}({\bm x}) = - 2\pi k_{\Tilde{\phi}} \delta^{(2)} ({\bm x}),\\
    &\Tilde{j_z} = -\frac{1}{g}\nabla^2\Tilde{s}^{\rm sol}({\bm x}) = 2\pi k_A\delta^{(2)} ({\bm x}). \label{eq:bosonic external sources}
\end{split}
\end{equation}
Here, $k_A= k_s/g$ is a numerical constant.
As is the case in the ANO string, a non-trivial configuration of ${\Tilde{\phi}}_r$ can be expressed by $J_{\Tilde{\phi}}$ regarded as the monopole charge of $-2\pi k_{{\Tilde{\phi}}}$.
Similarly, a non-trivial configuration of the gauge field $\Tilde{A}_z$ is sourced by the static magnetic current flowing along the string.
One should note that $k_A$ can be expressed as $k_A= 2\pi \widetilde{J}_{\rm tot}$, as is evident from the Gauss's law.
Therefore, unlike other numerical constants, $k_\phi,~k_e$ and $k_{\Tilde{\phi}}$, there is a strong restriction on $k_A$ due to the current quenching effect at least for the magnetic current.

\section{Cosmic string interactions: Analytic studies}\label{sec:interaction}

In this section, we investigate the interaction energy between two strings using the method proposed in Ref.~\cite{Speight:1996px}.
In the following analysis, we assume that two strings are straight and static, separated by a fixed distance, and that each string possesses a single winding number, $n=\pm 1$.
We will argue that the most dominant contribution comes from the lightest field in the underlying theory.

\subsection{Local strings interactions}\label{sec:local string interaction}

In this subsection, we compute interaction energies of local(ANO) strings.
When two strings are widely separated, effective descriptions of ANO strings with external sources in Eq.~\eqref{eq:total L1} are applicable.
We assume the whole field configuration can be approximated by the superposition of each source,
\begin{equation}
\begin{split}	
&J_\sigma = -2\pi k_{\phi1}\delta^{(2)} (\bm{x}-\bm{x}_1)-2\pi k_{\phi2}\delta^{(2)} (\bm{x-x}_2),~~~\\
&j_i=2\pi \frac{k_{e1}}{m_e} \epsilon_{ji} \nabla_j \delta^{(2)} (\bm{x-x}_1)+2\pi \frac{k_{e2}}{m_e}  \epsilon_{ji} \nabla_j \delta^{(2)} (\bm{x-x}_2) .\label{eq:dipole sources}
\end{split}
\end{equation}
Here, we introduce the two-dimensional Cartesian coordinate ${\bm x}=(x^1,x^2)$ which parameterizes the plane perpendicular to the string axes.
${\bm x}_1$ and ${\bm x}_2$ are positions of each string axe fixed by hand, while $k_{\phi i}$ and $k_{e i}$ ($i=1, 2$) are charges of each string.
The effective Lagrangian density of this system is defined by Eq.~\eqref{eq:total L1} with the above external sources.

The equations of motion obtained from Lagrangian density in Eq.~\eqref{eq:total L1} are given by
\begin{align}
(\Box + m_\phi^2)\sigma = - J_\sigma,~~~(\Box +m_e^2 ) U_\mu =  j_\mu. \label{eq:linearized eom}
\end{align}
One can solve the above differential equations by the standard Green's function method:
\begin{equation}
\begin{split}	
&\sigma^{\rm sol}(x) = \int \mathrm{d}^4 x'\,G(x-x';m_\phi)  J_\sigma(x'),\\
&U^{\rm sol}_{\mu} (x) =-\int \mathrm{d}^4 x'\,G(x-x';m_e) j_\mu(x') ,\label{eq:solutions}\\
&G(x-x',M)\equiv \int \frac{d^4 p}{(2\pi)^4} \frac{1}{p^2-M^2}e^{ip(x-x')}\,.
\end{split}
\end{equation}
Putting above solutions into original Lagrangian given by Eq.~\eqref{eq:total L1}, we obtain
\begin{align}
\mathcal{L} = -\frac{1}{2}J_\sigma \sigma^{\rm sol} -\frac{1}{2} U^{\rm sol}_\mu j^\mu.
\end{align}

The interaction energy is then evaluated as
\begin{align}
E_{\rm int} =  \frac{1}{2} \int dz \int d^2 x \int d^2 x' \left(-J_\sigma (x)D^{(2)}_\sigma (x-x') J_\sigma (x') + J_\mu (x) D_U^{(2)\mu\nu} (x-x') J_\nu (x') \right)\label{eq:formula}.
\end{align}
In this calculation, we have used the fact that 
$D^{(2)}_\sigma (x-x')$ and $D^{(2)\mu\nu}_U (x-x')$ are the two-dimensional Euclidean propagators defined by 
\begin{align}
&D^{(2)}_\sigma (x-x')=\int \frac{d^2 {\bm p}}{(2\pi)^2}\frac{1}{{\bm p}^2+m_\phi^2}e^{i {\bm p} \cdot ({\bm x}-{\bm x}')} =\frac{1}{2\pi}K_0 (m_\phi |{\bm x}-{\bm x}'|),\\
&D^{(2)\mu\nu}_U (x-x') =\frac{\eta^{\mu\nu}}{2\pi} K_0 (m_e |{\bm x} -{\bm x}'|).
\end{align}
Essentially, $K_0 (M|{\bm x}-{\bm x}'|)$ is the two-dimensional Yukawa potential with mass $M$.
Substituting the superposed external sources in Eq. \eqref{eq:dipole sources} into Eq.~\eqref{eq:formula}, we find 
\begin{align}
E_{\rm int} = \int dz 2\pi \left[k_{e1} k_{e2} K_0(m_e d)-k_{\phi1} k_{\phi2} K_0 (m_\phi d) \right],~d\equiv |{\boldsymbol x}_1-\boldsymbol{x}_2|. \label{eq:local interaction energy}
\end{align}
This result is in agreement with the previous studies~\cite{Bettencourt:1994kf,Speight:1996px}.

The first term of Eq.~\eqref{eq:local interaction energy} represents the interaction energy coming from the massive gauge field associated with the presence of the magnetic flux of the ANO strings, while the second term is the contribution from the massive scalar field.
The signs of Eq.~\eqref{eq:local interaction energy} simply reflect the fact that the gauge interaction through the magnetic flux is repulsive for winding numbers with the same signs, while the scalar interaction is always attractive since the charge of scalar field should always take the same sign.
Due to the asymptotic forms of $K_0(x)$, the interaction energy is exponentially suppressed on length scales longer than the inverse mass of the underlying field.
For $2e^2 / \lambda > 1$ ($m_\phi < m_e$), the scalar attraction is always dominant at $d\to \infty$.
Conversely, if $2e^2 / \lambda < 1$ ($m_\phi > m_e$), the repulsion sourced by the gauge field is dominant at $d\to \infty$ if the winding number of two strings are opposite.
These arguments are essentially independent of the precise values of the charges, $k_\phi$ and $k_e$, as long as they are finite values.
Therefore, the interaction energy of the string at $d \to \infty$ can be easily revealed by applying the point source formalism.

Once the values of $\lambda_\phi$ and $e^2$ are specified, $k_\phi$ and $k_e$ can be calculated explicitly as explained in Sec.~\ref{sec:local strings}.
In this case, as in the original analysis~\cite{Speight:1996px}, we can discuss which interactions are dominant at any $d$ .
We will do this in the next section and discuss the validity of the point source formalism by
comparing the results obtained by the non-linear numerical calculations with the analytic ones.

\subsection{Global strings interactions}\label{sec:global string interactions}

In this subsection, we estimate the interaction energy of two straight and static global strings.
In the following discussion, we first focus on the interaction energy from the topological charge.
The effective Lagrangian density of the NG-boson is given by Eq.~\eqref{eq:topological coupling} through its dual fields.
If the two global strings are far apart, the external sources can be approximated by a superposition of their respective sources,
\begin{align}
 J^{03}_B = -J^{30}_B = 2\pi n_1\eta_\phi\delta^{(2)} ({\bm x}-{\bm x}_1) + 2\pi n_2\eta_\phi \delta^{(2)} ({\bm x}-{\bm x}_2) \label{eq:massless sources},
\end{align}
where $n_1$ and $n_2$ are winding numbers of global strings whose axes placed at ${\bm x}={\bm x}_1$ and ${\bm x}={\bm x}_2$, respectively.
In the region sufficiently far from the string axis, one can use the approximation $|\phi|\simeq \eta_\phi $.
Then one can solve the equation of motion of the dual field $B_{\mu\nu}$ in Eq.~\eqref{eq:eom dual B} by using the standard Green's function method as
\begin{align}
B^{\mu\nu} (x) = \int \mathrm{d}^4 x'\, G(x-x',\epsilon) J_B^{\mu\nu} (x').
\end{align} 
Here, we fix a gauge by imposing $\partial_\mu B^{\mu\nu}=0$, which is analogous to the Coulomb gauge in the classical electrodynamics.\footnote{For a detailed discussion of the gauge degrees of freedom and its fixing for $B_{\mu\nu}$ field, see e.g. Ref.~\cite{Maluf:2018jwc}.}
We introduce infinitesimal mass $\epsilon$ of $B_{\mu\nu}$ to avoid infrared divergence, which will be taken to be zero at the end of calculation.
 
Then the interaction energy becomes
\begin{align}
E_{\rm int}[B]= \frac{1}{2} \int dz \int d^2 x \int d^2 x' \left( J_B^{\mu\nu} (x) D_B^{(2)} (x-x') J_{B\mu\nu} (x') \right)\label{eq:global string interactions},
\end{align}
where $D^{(2)}_{B}$ is defined by two-dimensional massless propagator,
\begin{align}
D^{(2)}_{B} (x-x') = \int  \frac{d^2 {\bm p}}{(2\pi)^2}\frac{1}{{\bm p}^2+\epsilon^2}e^{i {\bm x} \cdot ({\bm x}-{\bm x}')}=\frac{1}{2\pi}K_0 (\epsilon|{\bm x}-{\bm x}'|).
\end{align}
The interaction energy is then evaluated as
\begin{align}
E_{\rm int}[B] =  \int dz 4\pi \eta^2 n_1 n_2 K_0 \left(\epsilon d\right) \simeq - \int dz 4\pi\eta^2 n_1 n_2 \ln\left(\epsilon d \right).
\end{align}
The approximation of modified Bessel function $K_0 (\epsilon d) \simeq  -\log(\epsilon d)$ for $\epsilon d \ll 1$ is used in the last equality.
We redefine the origin of the interaction energy at $d=m_\eta ^{-1}$. This yields
\begin{align}
E_{\rm int}[B] \equiv  -\int dz 4\pi \eta^2 n_1 n_2 \left(\ln\left(\epsilon d\right)- \ln\left(\epsilon m_\eta^{-1}\right)\right)= -\int dz 4\pi \eta^2 n_1 n_2 \ln\left(m_\eta d\right).\label{eq:NG-boson interaction energy}
\end{align}
The above logarithmic dependence at long distance, $d\to \infty$, is in agreement with previous studies~\cite{Shellard:1987bv,Bettencourt:1994kf}.
The massless two-index antisymmetric tensor field leads to the long-range force.
The same-sign topological charges of two global strings can be regarded as the same magnetic charges of axions which lead to the repulsive force.

Let us next estimate the interaction energy of the massive mode.
Assuming that two strings are widely separated such that external sources can be approximated by the following form,
\begin{align}
J_G=\sqrt{2}\eta_\phi \left(\frac{n_1^2}{({\bm x}-{\bm x}_1)^2} + \frac{n_2^2}{({\bm x}-{\bm x}_2)^2}\right).
\end{align}
Then we obtain the interaction energy of $\sigma$ field  as
\begin{align}
E_{\rm int}[\sigma] = - \int dz \int d^2x \frac{2}{\lambda_\phi}\frac{n_1^2}{({\bm x}-{\bm x}_1)^2} \frac{n_2^2}{ ({\bm x}-{\bm x}_2)^2} . 
\end{align}
There exists divergent contribution at two string axes, ${\bm x}={\bm x}_i$ ($i=1,2$).
Obviously, effective descriptions by the external sources cannot be justified at these points.
Therefore we introduce the cutoff which is of the same order of the string thickness $\delta=1/(\lambda_\phi \sqrt{\eta_\phi})$.
Then one can evaluate the interaction energy as
\begin{align}
E_{\rm int}[\sigma] \simeq  - \int {\rm d}z\frac{2\pi}{\lambda_\phi}n_1^2 n_2^2 \frac{1}{d^2} \ln \left(\frac{\delta^2}{d^2}\right),~d\equiv |{\boldsymbol x}_1-\boldsymbol{x}_2|.
\end{align}
This interaction energy is suppressed by $1/d^2$ in the long-distance limit.
Thus, the interaction energy is dominated by the external source associated with the topological charge in \eqref{eq:NG-boson interaction energy} at $d\to \infty$.

\subsection{Bosonic superconducting strings interactions}

In this subsection, we compute the interaction energy of bosonic superconducting strings.
As discussed in Sec.~\ref{sec:bosonic superconducting}, there are four external sources corresponding to non-trivial field configurations in the $U(1)\times \widetilde{U}(1)$ model.
Since the asymptotic behavior of the string configurations associated with the spontaneously broken $U(1)$ is exactly the same as for the ANO string, the interaction energy originated from these fields is given by Eq.~\eqref{eq:local interaction energy}.
Here, we estimate the interaction energy sourced by the $\widetilde{U}(1)$ Higgs field, and the current trapped inside the string.
In the following discussion, we omit the contributions from the $U(1)$ sector, but they are always implicitly present.
An effective description of the widely separated bosonic superconducting strings is given by Eq.~\eqref{eq:sc interaction Lagrangian} with the following superimposed external sources,
\begin{equation}
\begin{split}
      J_{\Tilde{\phi}} = -2\pi k_{{\Tilde{\phi}}1}\delta^{(2)}({\bm x}-{{\bm x}}_1)-2\pi k_{{\Tilde{\phi}}2}\delta^{(2)}({\bm x}-{\bm x}_2),\\
      \Tilde{j_z} = -2\pi k_{A1} \delta^{(2)}({\bm x}-{\bm x}_1)-2\pi k_{A2} \delta^{(2)}({\bm x}-{\bm x}_2).
      \end{split}
\end{equation}
Calculation of the interaction energy is straightforward using the Green's function method. (See e.g. Sec.~\ref{sec:local string interaction} for detailed computations.)
The resultant interaction energy from $\widetilde{U}(1)$ sector is expressed as
\begin{align}
    E_{int}[{\Tilde{\phi}}, \Tilde{s}] = \int dz2\pi\left[-k_{{\Tilde{\phi}}1}k_{{\Tilde{\phi}}2}K_0(m_{\Tilde{\phi},\infty} d) + k_{A1}k_{A2}\ln \left( \frac{d}{\delta}\right)\right],~~~d = |{\bm x}_1-{\bm x}_2|.\label{eq:sc interaction energy}
\end{align}
Here, $\delta$ is the typical width of the strings.
The first term is originated from the localized condensate scalar field ${\Tilde{\phi}}$, while the second term comes from the current trapped inside the string.
The constant $k_{\Tilde{\phi}}$ does not depend on the sign of the winding number or the direction of the current.
Regardless of the directions of the two strings, $k_{\Tilde{\phi}1}$ and $k_{\Tilde{\phi}2}$ should take the same signature.
This implies that the first term is always attractive.
The second term, on the other hand, depends on the direction of the static currents.
If the direction of each current is the same (opposite), then it is an attractive (repulsive) force.

It is worth noting that the first term and the contribution from the $\widetilde{U}(1)$ sector are exponentially suppressed on their mass scales, whereas the second logarithmic term is always dominant at infinity.
This stems from the fact that the mass of ${\Tilde{\phi}}_r$ is generically non-vanishing at infinity, while the $\widetilde{U}(1)$ gauge field is massless because the condensate of ${\Tilde{\phi}}_r$ is localized inside the string.

\begin{figure}[t]
\centering\includegraphics[width=15cm]{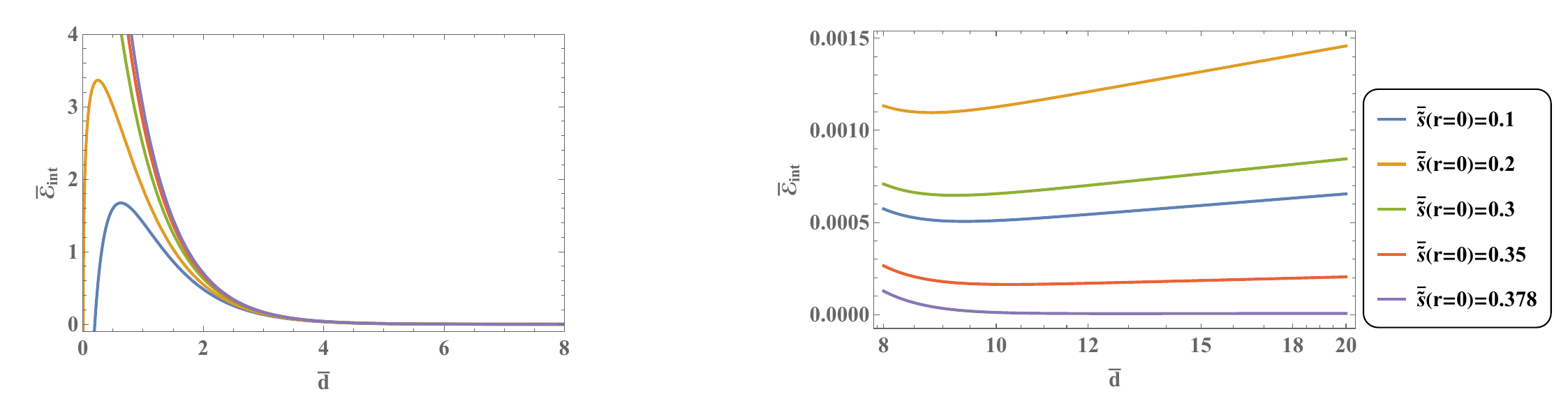}
\caption{The dependence of the interaction energy of bosonic superconducting cosmic strings on the separation distance $\Bar{d}$ is shown in the region of $0<\Bar{d}<8$ (left panel). The point source formalism for benchmark point A given in Table \ref{table:charge fitting} is used.
Also shown is a plot of $\Bar{\mathcal{E}}_{\rm int}$ in the region of $8<\Bar{d}<20$ using the same benchmark point.}
\label{fig:Speight interaction energy}
\end{figure}

Figure~\ref{fig:Speight interaction energy} shows the dependence of the interaction energy of bosonic superconducting cosmic strings, the sum of Eq.~\eqref{eq:local interaction energy} and Eq.~\eqref{eq:sc interaction energy}, on the separation distance for the benchmark point A specified in the App.~\ref{appendix:numerical calculation}. 
We used the normalized quantity $\Bar{d}$ and $\Bar{\mathcal{E}}_{int}$ in the figure according to the App.~\ref{appendix:numerical calculation}.
The charges $k_{\phi},~k_e,~k_{\tilde{\phi}}$  and $k_A$ are shown in Table \ref{table:charge fitting}.
As shown in the right panel of the figure, the logarithmic dependence of $\Bar{\mathcal{E}}_{int}$ on $\Bar{d}$ is observed in the long-distance limit.
$k_A$ is the charge associated with the $\widetilde{U}(1)$ current trapped inside the string, and is found to be one or two orders of magnitude smaller than the other charges due to current quenching effects, as explained in Sec.~\ref{sec:bosonic superconducting}.
Tiny values of $k_A$ have a small effect on $E_{\rm int}$ compared to other contributions when $d$ is comparable to or shorter than the mass scale of all fields in the underlying theory.
Thus, the effect of non-zero currents is only important in the long-distance limit.
This is a generic feature of bosonic superconducting strings and can also be observed for other parameter choices.

\section{Cosmic string interactions: Numerical studies}\label{sec:numerical study}

In this section, the interaction energy of static and straight strings is evaluated numerically for arbitrary separation distances.
In particular, the gradient flow method is used.
This method was recently employed in Ref.~\cite{Eto:2022hyt} to reveal the interaction energy of two conventional local strings and two local strings with Coleman-Weinberg potential.

The action for numerical simulations is the same as Eq.~(\ref{eq:superconducting lagrangian density}):
\begin{equation}
\begin{split}
    &S = \int {\rm d}^4x\left[-\frac{1}{4}F^{\mu\nu}F_{\mu\nu} + |D_\mu \phi|^2 -\frac{1}{4}\widetilde{F}^{\mu\nu}\widetilde{F}_{\mu\nu} +|\widetilde{D}_\mu \Tilde{\phi}|^2 - V(\phi,\Tilde{\phi})\right],\\
    &V(\phi,\Tilde{\phi}) = \frac{\lambda_\phi}{4} \left( |\phi|^2-\eta_\phi^2 \right)^2 + \frac{\lambda_{\Tilde{\phi}}}{4}\left(|\Tilde{\phi}|^2-\eta_{\Tilde{\phi}}^2\right)^2+\beta |\phi|^2 |\Tilde{\phi}|^2 .
\end{split}
\end{equation}
If one neglects the part of $\Tilde{\phi}$ and $\Tilde{A}_\mu$ with $\beta = 0$ and $g=0$, this action represents local strings, and if one further neglects the part of $A_\mu$ with $e=0$, it represents global strings. 

We use the normalization described in App.~\ref{appendix:local normalization} for local strings and superconducting strings. 
For global strings, we use another normalization described in App.~\ref{appendix:global normalization}.
We prepare a two-dimensional lattice plane and place
the two axes of the strings at $(\bar{x},\bar{y})=(\pm \bar{d}/2,0)$, where $(\bar{x},\bar{y})$ is the two-dimensional Cartesian coordinate.
Here, the position of string axis is defined as the point where the value of the scalar field, $\bar{\phi}$, vanishes, that is, $|\bar{\phi}|=0$.
In the following analysis, we will focus on the case where two strings having the unity winding numbers with the same signs.
The results for two strings with opposite winding numbers are discussed with point source formalism in Chap.~\ref{sec:interaction}.
Analysis of the interaction energy between cosmic strings with winding numbers more than unity is beyond the scope of the present paper.
The number of lattice sites is $600\times 600$ and the box length is $\bar{L} =30$ (corresponding to lattice spacing with $\bar{a}=0.05$), unless otherwise stated.
The details of the simulation setup and the numerical recipe are explained in App.~\ref{appendix:interaction energy}.

Before presenting the numerical results, let us explain how the axes of the two strings are fixed in the numerical simulation. 
If we do not fix the angular mode of $\bar{\phi}$ but fix the string axes by simply imposing $|\bar{\phi}|=0$ at $(\bar{x},\bar{y})=(\pm \bar{d}/2,0)$, the numerical simulation shows that four points of $\bar{\phi} \simeq 0$ are observed at large values of $\bar{\lambda}_{\phi}\gg 2$. 
Two of these points are located on the initial string axes $(\bar{x},\bar{y})=(\pm \bar{d}/2,0)$, but the other two should not appear.
Therefore, only imposing the condition $|\bar{\phi}|=0$ on the initial string axes may not guarantee a fixed separation distance $\bar{d}$. We thus fix the angular dependence of $\bar{\phi}$ in the whole region of the simulation box, in addition to the condition $|\bar{\phi}|=0$ at $(\bar{x},\bar{y})=(\pm \bar{d}/2,0)$ as in Ref.~\cite{Jacobs:1978ch}.
In this case, we confirm that the positions of the two string axes are maintained in the numerical simulation, and hence, the fixed separation distance $\bar{d}$ is well-defined.
(See App. \ref{appendix:interaction energy} for detailed procedures to fix the angular dependence of $\bar{\phi}$.)
For this reason, we fix the angular dependence of $\bar{\phi}$ for the entire simulation box in the following analysis.

\subsection{Local strings}

In this subsection, we present numerical results for the interaction energy of two local strings, originally calculated using the variational method in Ref.~\cite{Jacobs:1978ch}.

\begin{figure}[t]
\centering\includegraphics[width=15cm]{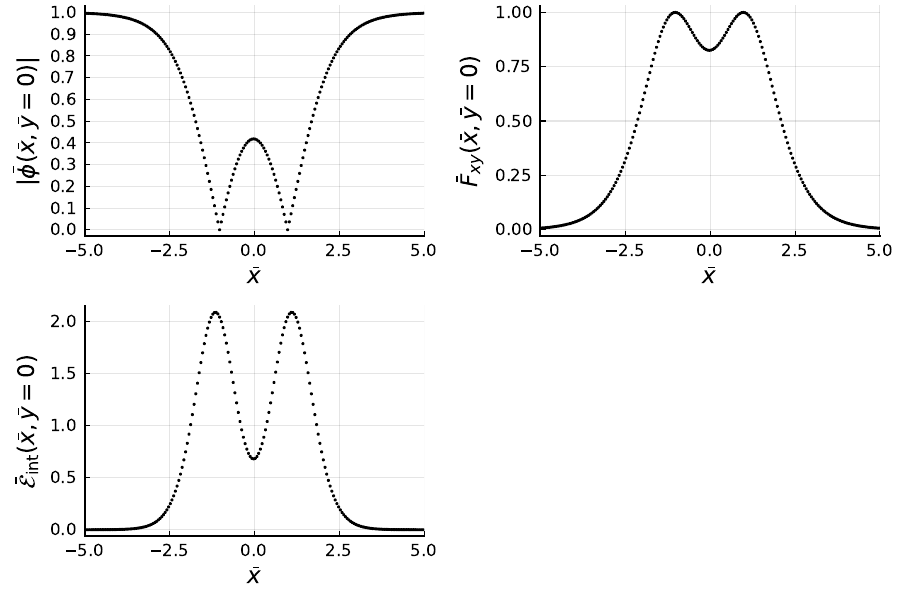}
\caption{
The scalar field magnitude $|\bar{\phi}(\bar{x},\bar{y}=0)|$ (top left), magnetic flux $\bar{F}_{xy}(\bar{x},\bar{y}=0)$ (top right), energy density $\bar{\mathcal{E}}_{\rm int}(\bar{x},\bar{y}=0)$ (bottom left) of two static local strings are shown for $\bar{d}=2$ with $\bar{\lambda}_\phi=2$.
}\label{fig:field configurations of two local strings}
\end{figure}
\begin{figure}[t]
\centering\includegraphics[width=15cm]{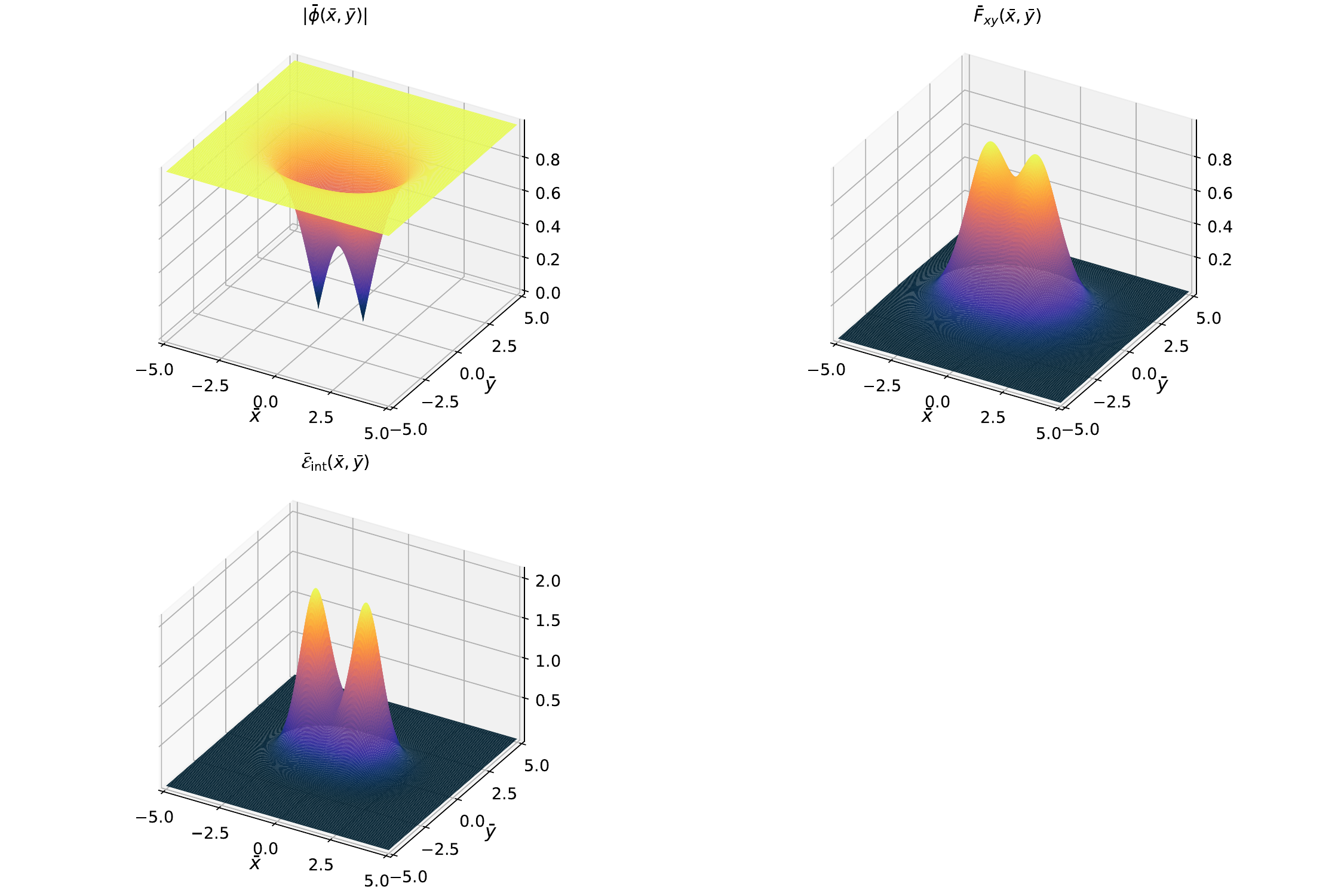}
\caption{
Surface plots of the scalar field magnitude $|\bar{\phi}(\bar{x},\bar{y})|$ (top left), magnetic flux $\bar{F}_{xy}(\bar{x},\bar{y})$ (top right), energy density $\bar{\mathcal{E}}_{\rm int}(\bar{x},\bar{y})$ (bottom left) of two static local strings are shown for $\bar{d}=2$ with $\bar{\lambda}_\phi =2$.}
\label{fig:field configurations of two local strings 2D}
\end{figure}

Figure~\ref{fig:field configurations of two local strings} shows the field configurations and the energy density of two local strings for $\bar{\lambda}_\phi = 2$ with a fixed separation, $\bar{d}=2$, on a $\bar{y}=0$ slice, while Fig.~\ref{fig:field configurations of two local strings 2D} displays these surface plots with the same parameter set.
We confirm that the configurations of $\bar{\phi}$ and $\bar{A}_i$ can be approximated by $\bar{\phi} = \bar{\phi}_+\bar{\phi}_-$ and $\bar{A}_i=\bar{A}_{i +}+\bar{A}_{i -}$, where $\bar{\phi}_\pm$ and $\bar{A}_{i \pm}$ ($i=1,2$) are the $n=1$ local string solutions whose axes are fixed at $(\bar{x},\bar{y})=(\pm \bar{d}/2,0)$ for sufficiently large $\bar{d}$.
It is also confirmed that the $n=2$ solution of local strings is reproduced in the limit of $\bar{d}=0$ as it should be.

\begin{figure}[t]
\centering\includegraphics[width=9cm]{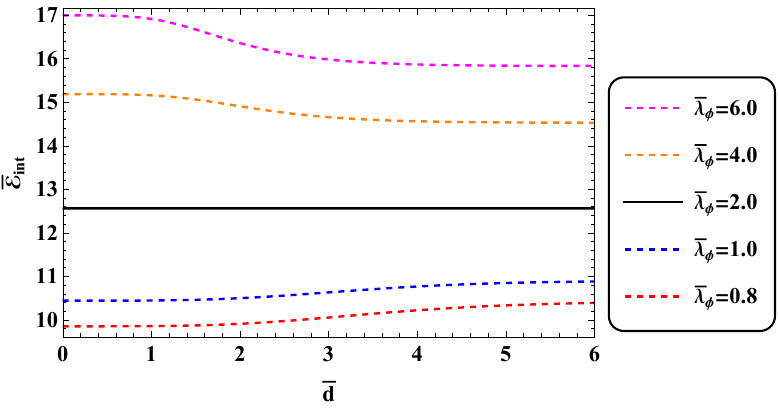}
\caption{
The dependence of the interaction energy density, $\bar{\mathcal{E}}_{\rm int}$, on the distance, $\bar{d}$, are shown for various couplings, $\bar{\lambda}_\phi$, in the case of local strings.
}\label{fig:local_string_int}
\end{figure}

Figure~\ref{fig:local_string_int} displays the dependence of the energy density on $\bar{d}$ at different values of $\bar{\lambda}_\phi$.
From this figure, it is easy to see that the dependence of $\bar{\mathcal{E}}_{\rm int}$ on $\bar{d}$ can be classified into three cases: $\bar{\lambda}_\phi \gtrless 2$, $\bar{\lambda}_\phi =2$.
In the special case with $\bar{\lambda}_\phi =2$, $\bar{\mathcal{E}}_{\rm int}$ does not depend on $\bar{d}$.
For $\bar{\lambda}_\phi> 2$, $\bar{\mathcal{E}}_{\rm int}$ becomes smaller as $\bar{d}$ is increased.
In this case, a stable solution is obtained in the long distance limit, $\bar{d}\to \infty$, but the $n=2$ ANO solution is unstable against perturbations of $\bar{d}$.
Conversely, for $\bar{\lambda}_\phi < 2$, $\bar{\mathcal{E}}_{\rm int}$ becomes smaller as $\bar{d}$ becomes smaller.
Therefore, the $n=2$ ANO solution is stable for $\bar{\lambda}_\phi < 2$ in contrast to the case with $\bar{\lambda}_\phi >2$. 
It should be noted that, since the field configuration with $n=2$ is a static solution to the equation of motion, the first derivative of $\bar{\mathcal{E}}_{\rm int}$ with respect to $\bar{d}$ must vanish at $\bar{d}=0$ for an arbitrary $\bar{\lambda}_\phi$.  
The stability of local strings is generically difficult to capture with an analytic approach when $\bar{d}$ is small, except for $\bar{\lambda}_\phi = 2.0$. 
However, as explicitly shown in Ref.~\cite{Eto:2022hyt}, analytic estimation is possible by using perturbations around the BPS state if $\bar{\lambda}_\phi$ is close to the BPS limit $\bar{\lambda}_\phi \simeq 2$. 

\begin{figure}[t]
\centering\includegraphics[width=7.5cm]{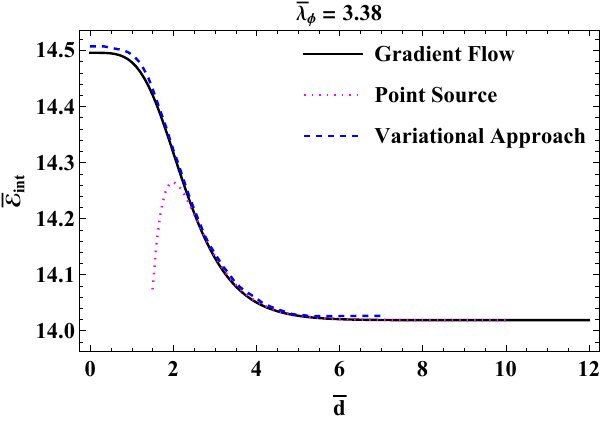}
\centering\includegraphics[width=7.5cm]{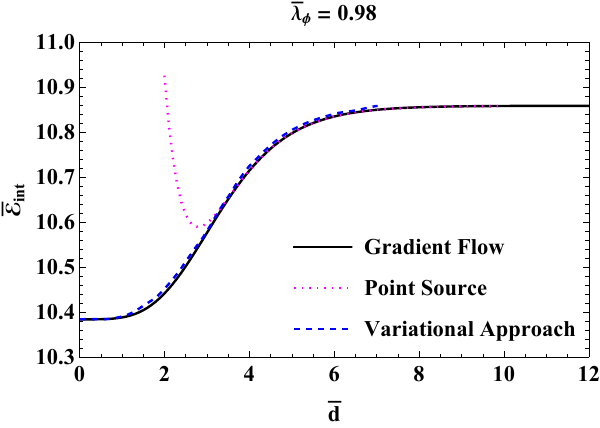}
\caption{
The left figure shows the results for $\bar{\lambda}_\phi=3.38$ and the right figure shows those for $\bar{\lambda}_\phi=0.98$.
The interaction energies estimated by the full numerical calculation using the gradient flow method (black curve), the point source formalism (magenta dotted curve), and the variational approach in Ref.~\cite{Jacobs:1978ch} (blue dashed curve) are shown.}
\label{fig:local_point}
\end{figure}

Finally, in Fig.~\ref{fig:local_point}, the results obtained by our full numerical calculation based on the gradient flow method are compared with those obtained by the point source formalism and the variational approach~\cite{Jacobs:1978ch}.
In the figure, we focus on two benchmark points used in Ref.~\cite{Jacobs:1978ch}.
We see that our results are in very good agreement with those of Ref.~\cite{Jacobs:1978ch}.
For short distances where the inverse of the lightest mass among the scalar and the gauge bosons is longer than $\bar{d}$, the results of the point source formalism deviate from the full numerical calculation.
In particular, analyses relying on the point source formalism indicate the existence of a saddle point or a minimum of $\bar{\mathcal{E}}_{\rm int}$ with respect to $\bar{d}$, suggesting a non-trivial phase structure of the local string. Our full numerical calculations, however, did not find such a non-trivial phase structure. This is because an effective description of the local string by the point source formalism cannot be justified for small $\bar{d}$.

\subsection{Global strings}\label{sec:global numerical}

In this subsection, we present numerical results for the interaction energy of two straight global strings. A numerical study of the interaction energy of two global strings is given in Ref.~\cite{Perivolaropoulos:1991du}.

\begin{figure}[t]
\centering\includegraphics[width=13cm]{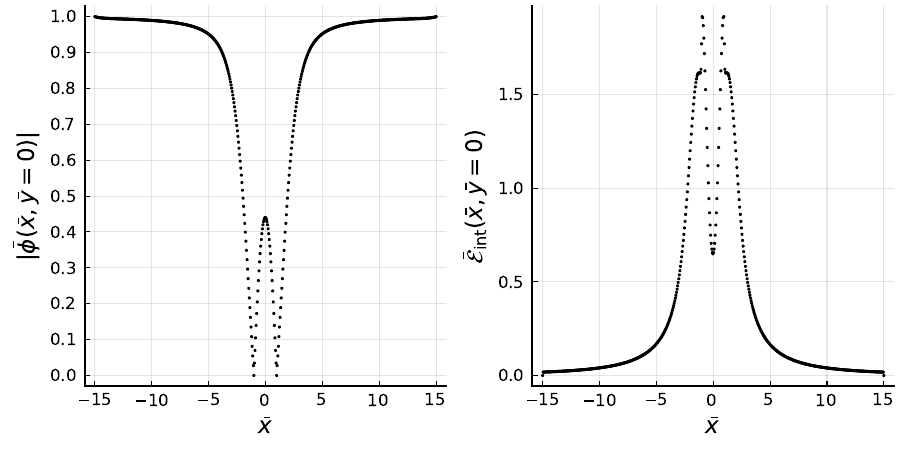}
\caption{
The scalar field magnitude $\bar{\phi}( \bar{x}, \bar{y}=0)$ (left) and energy density $\bar{\mathcal{E}}_{\rm int}( \bar{x}, \bar{y}=0)$ (right) for two static global strings with $\bar{d}=2$ and $\bar{\lambda}_\phi=4$ are shown.}
\label{fig:field configurations of global strings}
\end{figure}
\begin{figure}[t]
\centering\includegraphics[width=15cm]{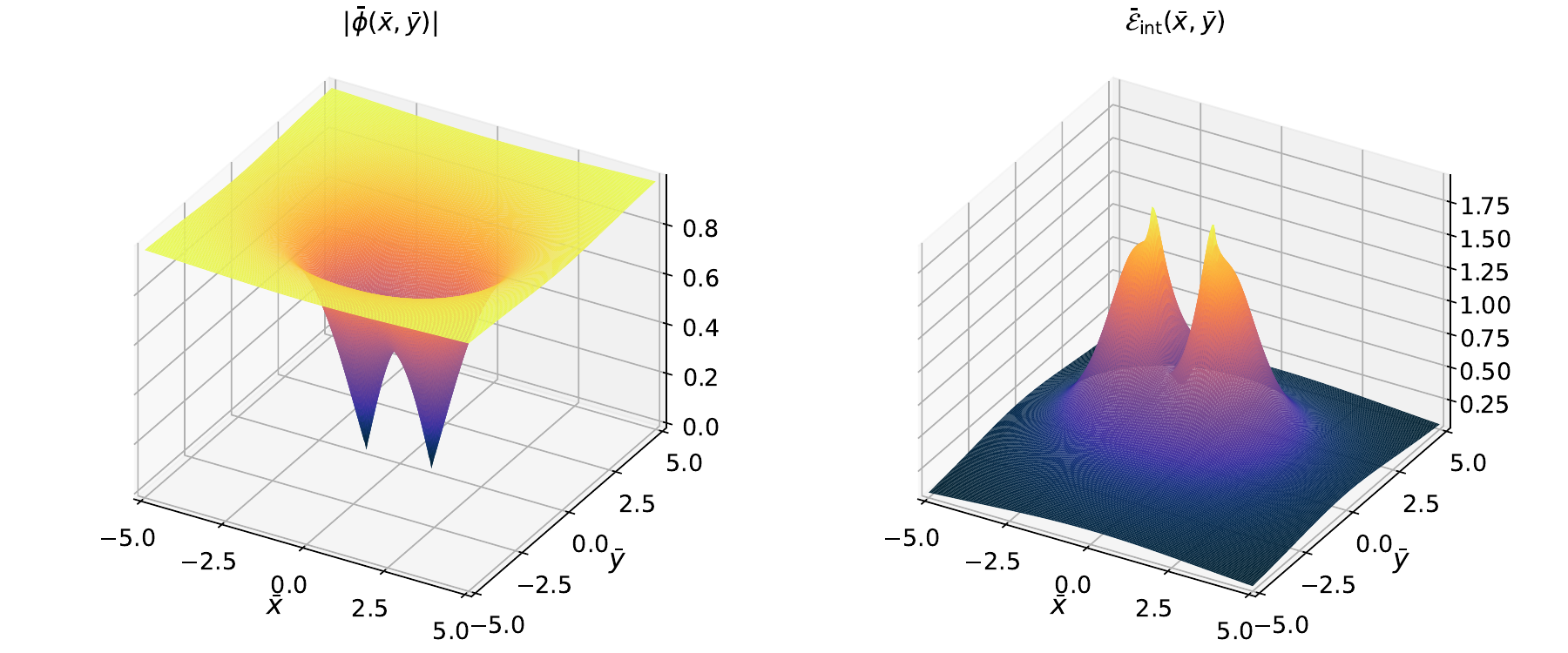}
\caption{
Surface plots of the scalar field magnitude $|\bar{\phi}(\bar{x},\bar{y})|$ (left), and the energy density $\bar{\mathcal{E}}_{\rm int}(\bar{x},\bar{y})$ (right) for two static global string configurations for $\bar{d}=2$ with $\bar{\lambda}_\phi =4$ are shown.
}\label{fig:field configurations of two global strings 2D}
\end{figure}

Figure~\ref{fig:field configurations of global strings} shows the field configuration of $|\bar{\phi}|$ and the energy density of two global strings on $\bar{y}=0$ slice for $\bar{\lambda}_\phi = 4$ with the fixed separation, $\bar{d}=2$, while Fig.~\ref{fig:field configurations of two global strings 2D} is a surface plot for the same parameter set.
We have numerically confirmed that the whole field configuration cannot be well approximated by the form of superposition ansatz, $\bar{\phi}=\bar{\phi}_+\bar{\phi}_-$, where $\bar{\phi}_{\pm}$ is the global string configuration with unit winding number. 
Their axes are placed at $(\bar{x},\bar{y})=(\pm \bar{d}/2,0)$, similarly to the local string case.
The interaction energy density $\bar{\mathcal{E}}_{\rm int}(\bar{x},\bar{y})$ has a kink around the string axes because the position of the string axes are fixed by hand.

\begin{figure}[t]
\centering
\includegraphics[width=7.5cm]{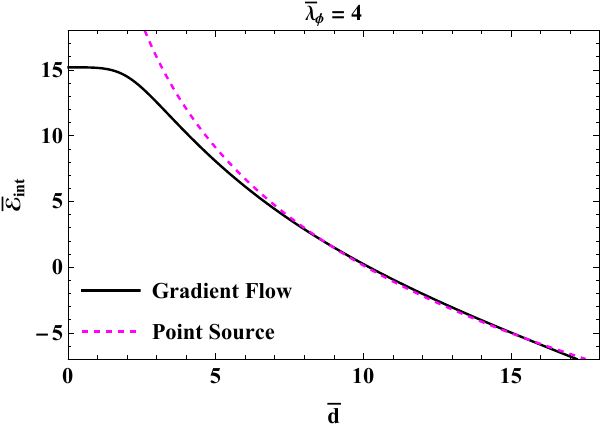}
\includegraphics[width=7.5cm]{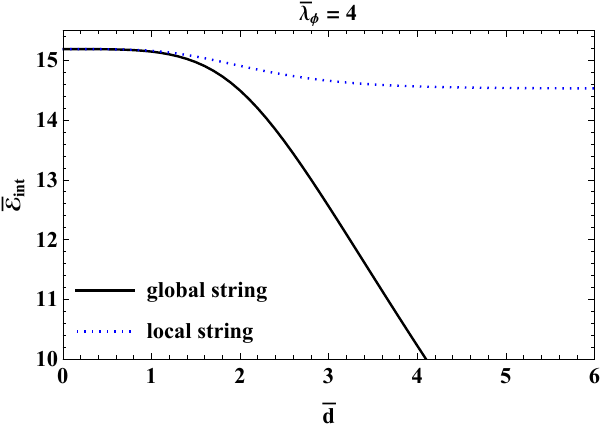}
\caption{
The dependence of interaction energy, $\bar{\mathcal{E}}_{\rm int}$, on the fixed distance, $\bar{d}$, is shown in the case of two global strings for $\bar{\lambda}_\phi=4$ (left panel). Also shown is the comparison with local strings with the same $\bar{\lambda}_\phi$ and $e=1$ (right panel).
}\label{fig:global_string_int}
\end{figure}

Figure~\ref{fig:global_string_int} shows the dependence of $\bar{\mathcal{E}}_{\rm int}$ on $\bar{d}$ in the case of a global string with $\bar{\lambda}_\phi =4$.
In the right panel of the figure, for clarity, the origin of the vacuum energy of the local string is shifted such that its $\bar{\mathcal{E}}_{\rm int}$ at $\bar{d}=0$ coincides with that of the global string.
The logarithmic dependence of $\bar{\mathcal{E}}_{\rm int}$ on $\bar{d}$ is observed at large $\bar{d}$, which is in good agreement with the results of the point source formalism and previous study~\cite{Perivolaropoulos:1991du}.
When the distance from the string axis is closer than $m_e^{-1}$, the angular mode of $\bar{\phi}$ is not canceled by the gauge field, so the behavior of the local string is almost identical to that of the global string.
Thus, when $d\lesssim m^{-1}_e$, the behavior of $\bar{\mathcal{E}}_{\rm int}$ of the local string is the same as that of the global string. 
This feature can be seen from the right panel of Fig.~\ref{fig:global_string_int}.
Therefore, the phase structure of the interaction energy of the global string can be understood as the limit of $e\to 0~(m_e^{-1}\to \infty)$.

\subsection{Bosonic superconducting strings without current}

In this subsection, we show our numerical results for the interaction energy of two bosonic superconducting strings without current.
It can be seen that the introduction of an additional scalar field whose condensate is localized inside the string has a dramatic effect on the interaction energy in a certain parameter region.

\begin{figure}[t]
\centering\includegraphics[width=15cm]{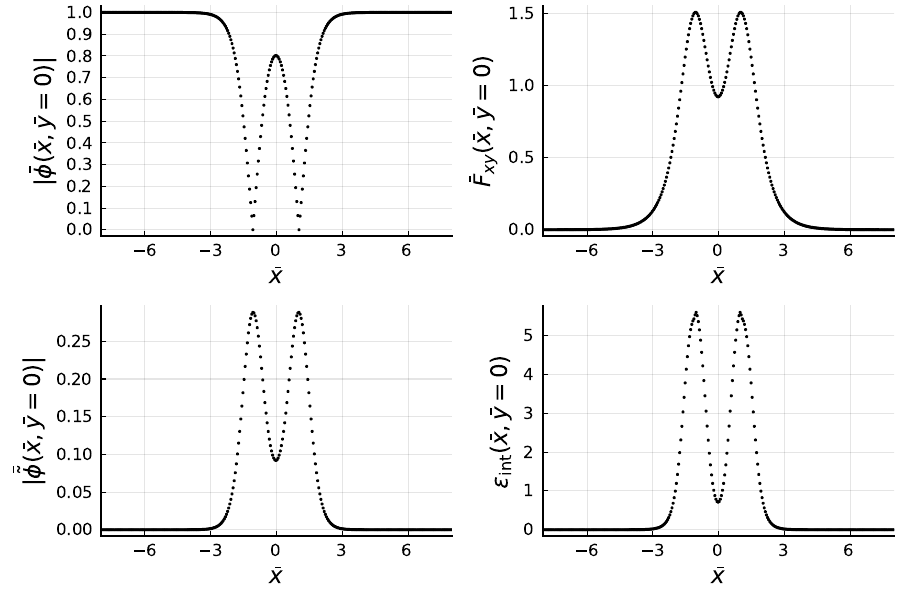}
\caption{
The scalar field magnitude $|\bar{\phi}(\bar{x},\bar{y}=0)|$ (top left), the magnetic flux $\bar{F}_{xy}(\bar{x},\bar{y}=0)$ (top right), the $|\bar{\Tilde{\phi}}(\bar{x},\bar{y}=0)|$ field (bottom left), and the energy density $\bar{\mathcal{E}}_{\rm int}(\bar{x},\bar{y}=0)$ (bottom right) of two static superconducting string configurations with no current are shown for $\bar{d}=2.1$ and $\bar{\lambda}_\phi = 8,~\bar{\lambda}_\sigma = 80,~\bar{\beta} = 24$ and $\bar{\eta}_{\Tilde{\phi}}=0.55$.
}\label{fig:field configurations of two superconducting strings without current}
\end{figure}
\begin{figure}[t]
\centering\includegraphics[width=15cm]{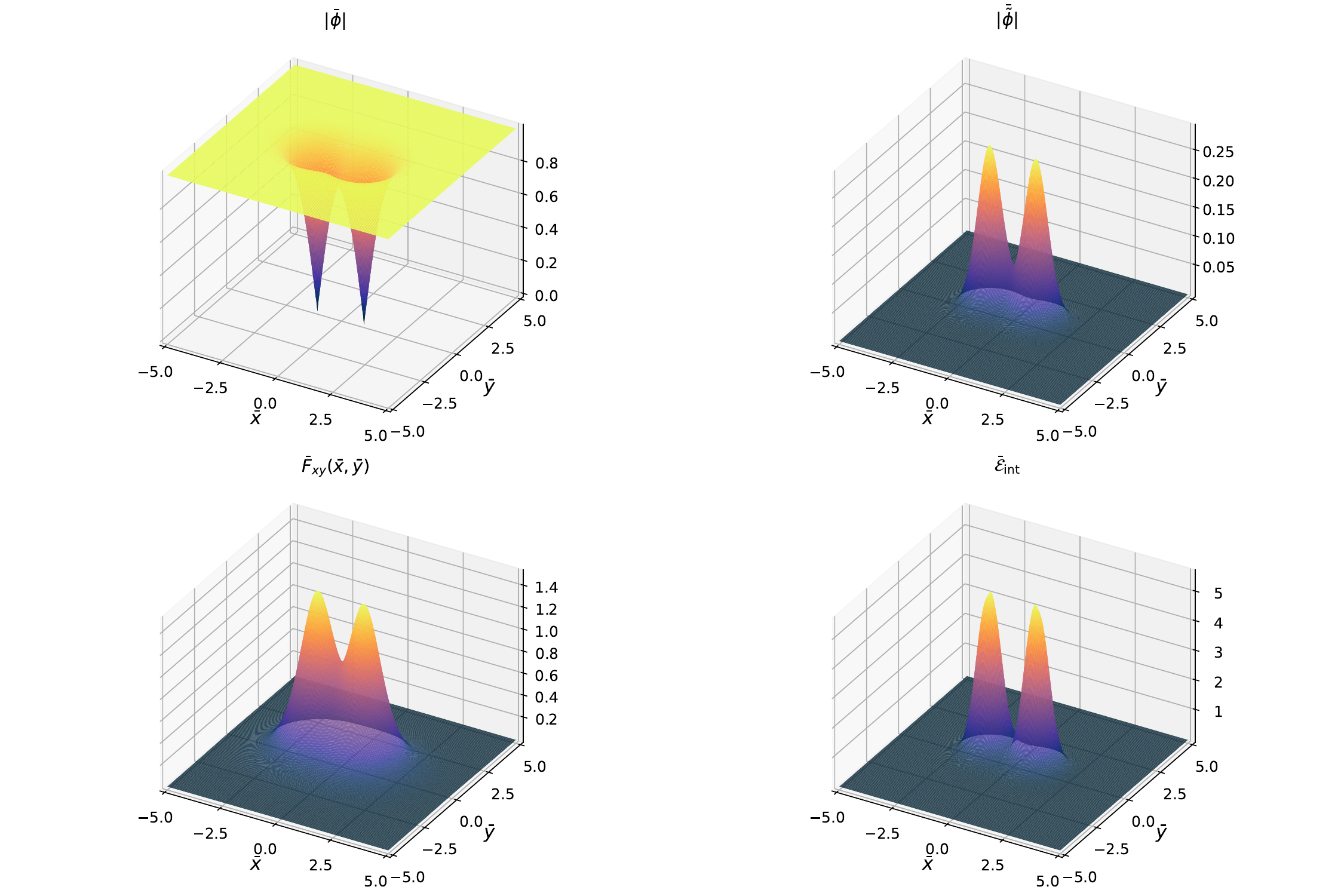}
\caption{
Surface plots of the scalar field magnitude $|\bar{\phi}(\bar{x},\bar{y})|$ (top left), the magnetic flux $\bar{F}_{xy}(\bar{x},\bar{y})$ (top right), the $|\bar{\Tilde{\phi}}(\bar{x},\bar{y})|$ field (bottom left), and the energy density $\bar{\mathcal{E}}_{\rm int}(\bar{x},\bar{y})$ (bottom right) of two static superconducting string configurations without current are shown for $\bar{d}=2.1$ with $\bar{\lambda}_\phi = 8,~\bar{\lambda}_{\Tilde{\phi}} = 80,~\bar{\beta} = 24$ and $\bar{\eta}_{\Tilde{\phi}}=0.55$.
}\label{fig:field configurations of two superconducting strings without current 2D}
\end{figure}

\begin{figure}[t]
\centering\includegraphics[width=15cm]{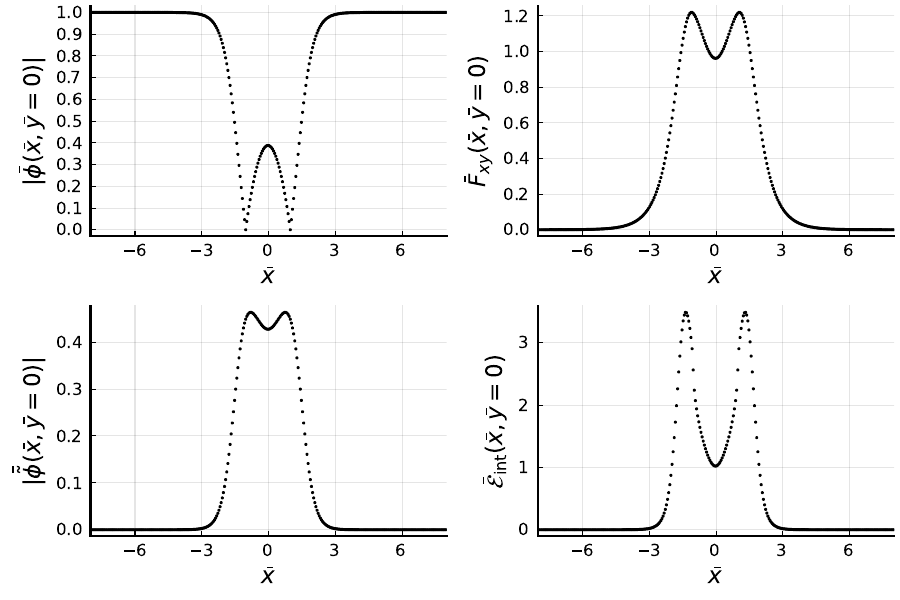}
\caption{
The scalar field magnitude $|\bar{\phi}(\bar{x},\bar{y}=0)|$ (top left), the magnetic flux $\bar{F}_{xy}(\bar{x},\bar{y}=0)$ (top right), the $|\bar{\Tilde{\phi}}(\bar{x},\bar{y}=0)|$ field (bottom left), and the energy density $\bar{\mathcal{E}}_{\rm int}(\bar{x},\bar{y}=0)$ (bottom right) of two static superconducting string configurations without current are shown for $\bar{d}=2$ with $\bar{\lambda}_\phi = 8,~\bar{\lambda}_\sigma = 80,~\bar{\beta} = 24$ and $\bar{\eta}_{\Tilde{\phi}}=0.55$.
}\label{fig:field configurations of two superconducting strings without current05}
\end{figure}

\begin{figure}[t]
\centering
\includegraphics[width=15cm]{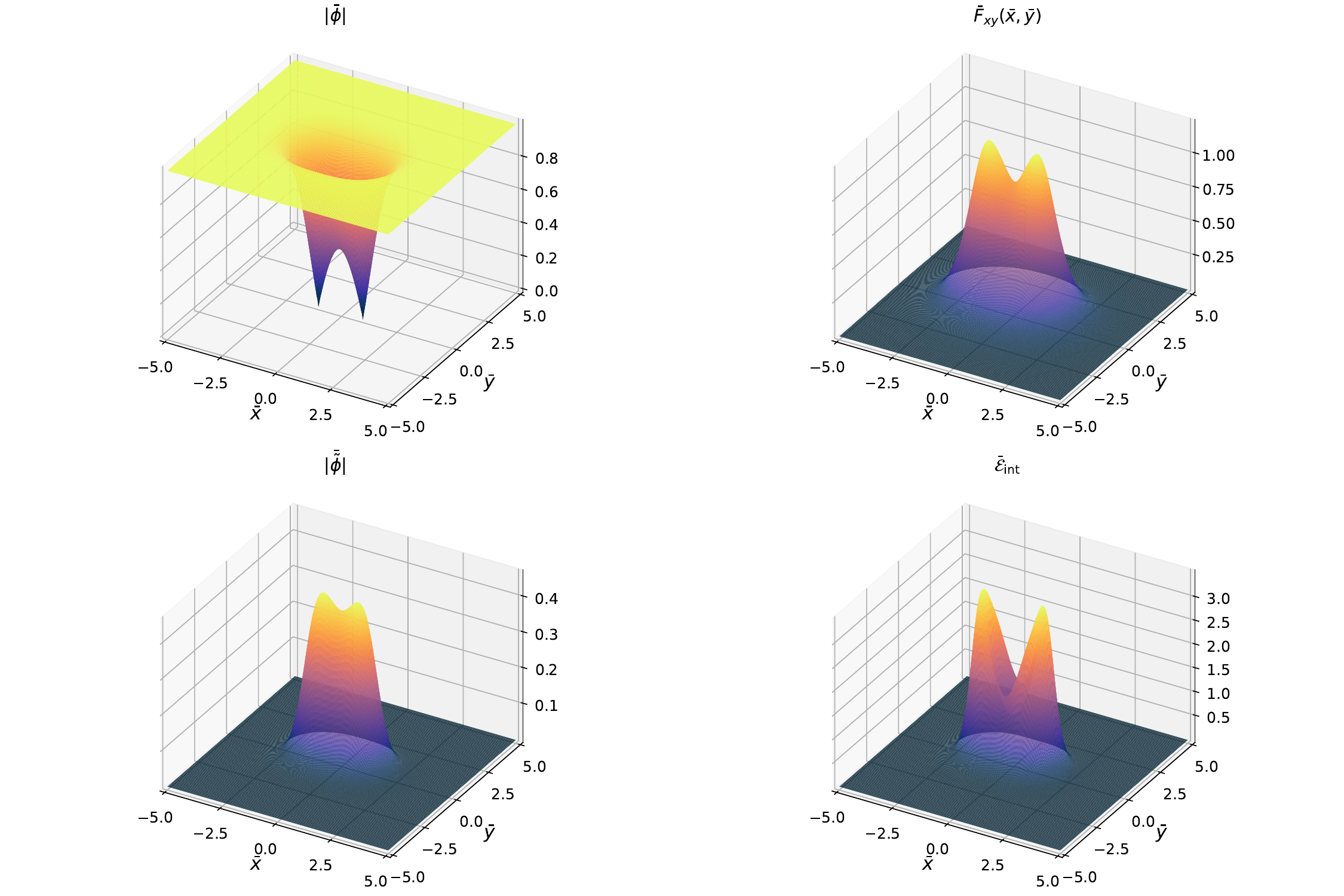}
\caption{
Surface plots of the scalar field magnitude $|\bar{\phi}(\bar{x},\bar{y})|$ (top left), the magnetic flux $\bar{F}_{xy}(\bar{x},\bar{y})$ (top right), the $|\bar{\Tilde{\phi}}(\bar{x},\bar{y})|$ field (bottom left), and the energy density $\bar{\mathcal{E}}_{\rm int}(\bar{x},\bar{y})$ (bottom right) of two static superconducting string configurations without current are shown for $\bar{d}=2$ with $\bar{\lambda}_\phi = 8,~\bar{\lambda}_{\Tilde{\phi}} = 80,~\bar{\beta} = 24$ and $\bar{\eta}_{\Tilde{\phi}}=0.55$.
}\label{fig:field configurations of two superconducting strings without current 2D05}
\end{figure}

Figure~\ref{fig:field configurations of two superconducting strings without current} shows field configurations and the energy density of two superconducting strings on $\bar{y}=0$ slice with fixed separation, $\bar{d}=2.1$, for the benchmark point with $\bar{\lambda}_\phi=8,~\bar{\lambda}_{\Tilde{\phi}}=80,~\bar{\beta} =24$ and $\bar{\eta}_{\Tilde{\phi}}=0.55$.
Figure~\ref{fig:field configurations of two superconducting strings without current 2D} is a surface plot for the same parameter set.
The whole field configurations can be approximated by the superposition ansatz, $\bar{\phi} = \bar{\phi}_+\bar{\phi}_-,~\bar{A}_i = \bar{A}_{i +}+\bar{A}_{i -}$, and $\bar{\Tilde{\phi}}_r= \bar{\Tilde{\phi}}_{r+}+\bar{\Tilde{\phi}}_{r-}$, where $\bar{\phi}_{\pm},~\bar{A}_{i \pm}$ and $\bar{\Tilde{\phi}}_{r\pm}$ are the configurations of each string whose axis is placed at $(\bar{x},\bar{y})=(\pm \bar{d}/2,0)$, respectively.
The same benchmark point is taken as in Figs.~\ref{fig:field configurations of two superconducting strings without current05} and \ref{fig:field configurations of two superconducting strings without current 2D05} for $\bar{d}=2$.
By comparing the condigurations at $\bar{d}=2.1$ and $\bar{d}=2$, we can see the coalescence of $\bar{\tilde{\phi}}$ happens at $\bar{d}=2$ where the magnitude of $\bar{\tilde{\phi}}$ enlarges and configurations of two string attach to each other.
This coalescence cannot be captured by the point source formalism and leads to a characteristic phase structure of interaction energy, as we will discuss below.


\begin{figure}[b]
\centering
\includegraphics[width=7.5cm]{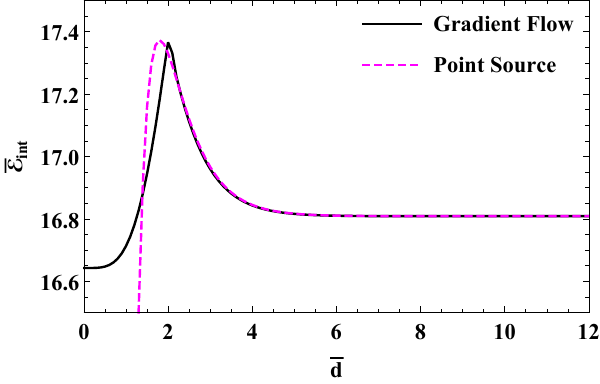}
\caption{
The dependence of the interaction energy, $\bar{\mathcal{E}}_{\rm int}$, on the separation distance, $\bar{d}$, is shown in the case of two bosonic superconducting strings without current for $\bar{\lambda}_{\phi} = 8,~\bar{\lambda}_{\Tilde{\phi}} = 80,~\bar{\beta} = 24$ and $\bar{\eta}_{\Tilde{\phi}}=0.55$.
}\label{fig:superconducting_int_with_point_source}
\end{figure}

\begin{figure}[t]
\includegraphics[width=7.5cm]{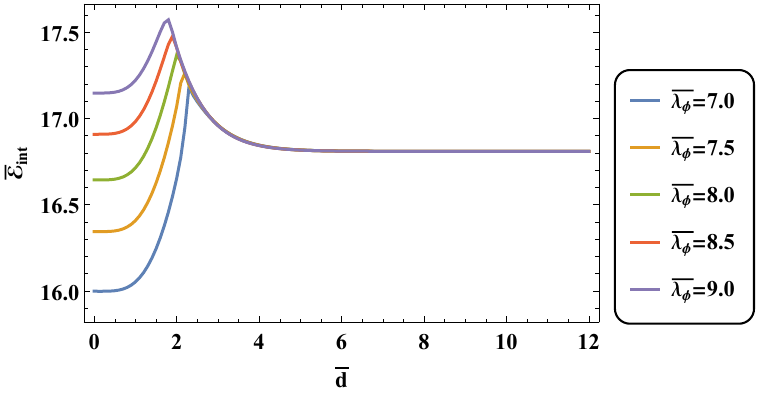}
\includegraphics[width=7.5cm]{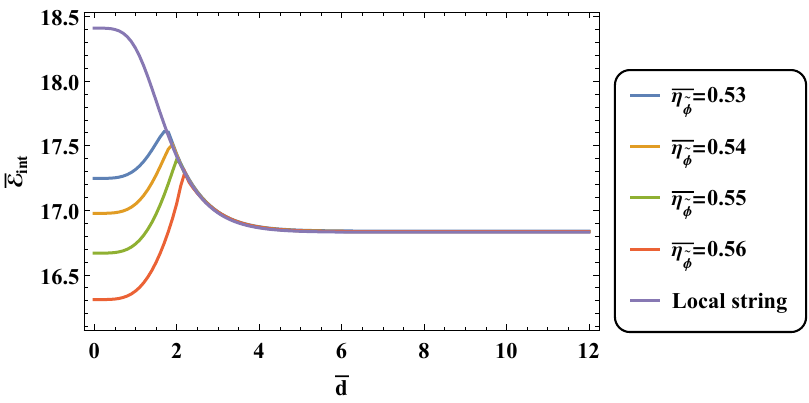}
\includegraphics[width=7.5cm]{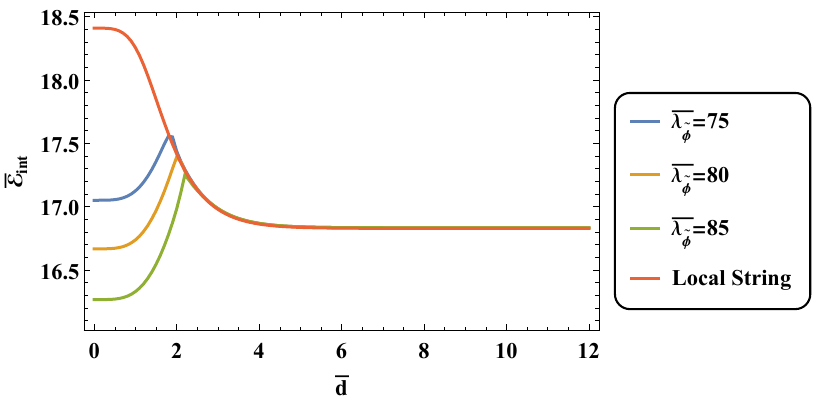}
\includegraphics[width=7.5cm]{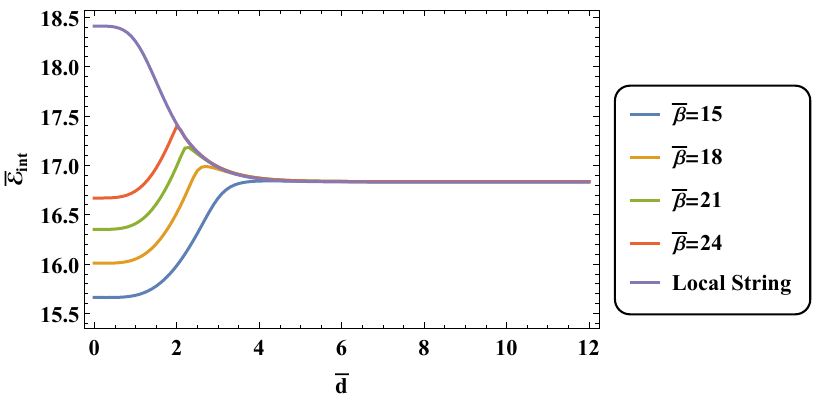}
\caption{
The dependence of interaction energies of bosonic superconducting strings, $\bar{\mathcal{E}}_{\rm int}$, on the fixed distance, $\bar{d}$, are shown for $\bar{\lambda}_{\Tilde{\phi}} = 80,~\bar{\beta} = 24,~\bar{\eta}_{\Tilde{\phi}}=0.55$ with various $\bar{\lambda}_{\phi}$ (top left), $\bar{\lambda}_\phi = 8,~\bar{\lambda}_{\Tilde{\phi}} = 80,~\bar{\beta} = 24,~\bar{\eta}_{\Tilde{\phi}}=0.55$ with various $\bar{\eta}_{\tilde{\phi}}$ (top right), $\bar{\lambda}_{\phi} = 8,~\bar{\beta} = 24,~\bar{\eta}_{\Tilde{\phi}}=0.55$ with various $\bar{\lambda}_{\Tilde{\phi}}$ (bottom left) and  $\bar{\lambda}_{\phi}=8$,~$\bar{\lambda}_{\Tilde{\phi}} = 80,~\bar{\eta}_{\Tilde{\phi}}=0.55$ with various $\bar{\lambda}_{\phi}$ (bottom right). The vacuum energy is unified with $\bar{\lambda}_\phi = 8,~\bar{\lambda}_{\Tilde{\phi}} = 80,~\bar{\beta} = 24,~\bar{\eta}_{\Tilde{\phi}}=0.55$ for illustrative purposes.
}\label{fig:superconducting_interaction energies_scan1}
\end{figure}

Figure~\ref{fig:superconducting_int_with_point_source} shows the dependence of $\bar{\mathcal{E}}_{\rm int}$ on $\bar{d}$ by the gradient flow method and by the point source formalism for the benchmark point with $\bar{\lambda}_{\phi}=8,~\bar{\lambda}_{\Tilde{\phi}}=80,~\bar{\beta}=24$ and $\bar{\eta}_{\Tilde{\phi}}=0.55$.
As can be seen from the figure, the results relying on the point source formalism are in good agreement with those of the gradient flow method at infinity.
Figure~\ref{fig:superconducting_interaction energies_scan1} shows the dependence of $\bar{\mathcal{E}}_{\rm int}$ on the separation distance $\bar{d}$ for various $\bar{\lambda}_\phi$, $\bar{\eta}_{\tilde{\phi}}$, $\bar{\lambda}_{\tilde{\phi}}$ and $\bar{\beta}$ in the parameter space which allows the formation of bosonic superconducting strings with the unit winding number.
In the figure, the origin of the vacuum energy is adjusted such that all interaction energies coincide at $\bar{d}\to \infty$ for visibility.
In order to understand the effect of $\bar{\tilde{\phi}}_r$ on $\bar{\mathcal{E}}_{\rm int}$, we compare $\bar{\mathcal{E}}_{\rm int}$ of the bosonic superconducting string with that of the local string in the figures.
These figures show that the dependence of $\bar{\mathcal{E}}_{\rm int}$ on $\bar{d}$ is almost the same as that of local strings.
On the other hand, it is drastically different for small $\bar{d}$ because $\bar{\mathcal{E}}_{\rm int}$ is a decreasing function with respect to $\bar{d}$.
A remarkable feature is that the first derivative of $\bar{\mathcal{E}}_{\rm int}$ looks discontinuous at the transition point, $\bar{d}=\bar{d}_c$, within the resolution of our lattice setup.
We suppose that this kink might be smoothened as we improve the lattice resolution.
We find that this kink behavior is triggered by the coalescence of $\bar{\tilde{\phi}}$ since the coalescence of $\bar{\tilde{\phi}}$ also happens at $\bar{d}=\bar{d}_c$.

The dependence of $\mathcal{E}_{\rm int}$ on $d$ is roughly understood as follows.
For bosonic superconducting strings (without current), there are three important length scales associated with the scalar field $|\phi|$, the $U(1)$ gauge field, and $\tilde{\phi}$, namely, $m_{\phi}, m_e$, and $m_{\tilde{\phi}}$ evaluated at infinity.
For all benchmark points taken in the Fig.~\ref{fig:superconducting_interaction energies_scan1}, there is a hierarchical structure $m_{e}^{-1}> m_{\phi}^{-1}>m_{\tilde{\phi}}^{-1}$.
On the length scale $d\gtrsim 2m_e^{-1}$, effective descriptions by the point source formalism can be applied because non-linear effects caused by all fields are clearly irrelevant.
For $ 2m_e^{-1}\gtrsim d\gtrsim 2m_{\phi}^{-1}$ or $ 2m_\phi^{-1}\gtrsim d\gtrsim 2m_{\tilde{\phi}}^{-1}$, the non-linear effects of the $U(1)$ sector are important, and hence, analysis relying on the point source formalism may not apply.
However, in this region, $\mathcal{E}_{\rm int}$ is almost the same as that of the local string because of the small effect of $\tilde{\phi}_r$.
This feature can be observed in Fig.~\ref{fig:superconducting_interaction energies_scan1}.
At $d=d_c\approx 2m^{-1}_{\tilde{\phi}}$, the aforementioned coalescence of $\tilde{\phi}_r$ takes place, and the non-linear effect of $\tilde{\phi}$ is important for $ 2m_{\Tilde{\phi}}^{-1}\gtrsim d$.
In this region, the configuration of $\tilde{\phi}_r$ possesses approximate circular symmetry, and the gradient energy of $\tilde{\phi}_r$ gives rise to the attraction.
This additional attraction due to $\tilde{\phi}_r$ gives the bosonic superconducting string a richer phase structure than the conventional local string.
The effects of $\eta_{\Tilde{\phi}},~\lambda_{\Tilde{\phi}},~\lambda_\phi$, and $\beta$ on the phase structure of the interaction energy may be qualitatively understood by these arguments since these parameters are related to the mass parameters, $m_e,~m_{\phi}$ and $m_{\Tilde{\phi}}$.
These qualitative features are numerically confirmed to be the same for other hierarchical mass structures.

\subsection{Bosonic superconducting strings with current}\label{sec:sc with current}

In this subsection, we discuss numerical results including the effect of non-zero currents associated with the $\widetilde{U}(1)$ gauge field.
Because the massless $\widetilde{U}(1)$ gauge field has a long-range effect on another string, we expand the lattice simulation box is from $600\times 600$ to $1000\times 1000$ to see long-distance behavior and to have enough space to decouple two strings while keeping the lattice space $\bar{a}=0.05$.

\begin{figure}[t]
\centering\includegraphics[width=15cm]{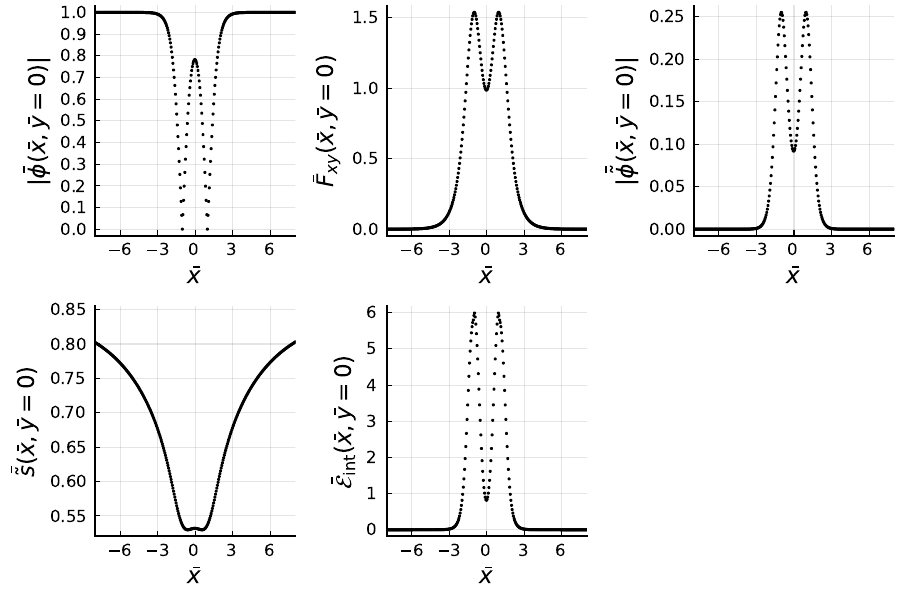}
\caption{
The scalar field magnitude $|\bar{\phi}(\bar{x},\bar{y}=0)|$ (top left), the magnetic flux $\bar{F}_{xy}(\bar{x},\bar{y}=0)$ (top middle), the $|\bar{\Tilde{\phi}}(\bar{x},\bar{y}=0)|$ field (top right), the $\bar{\Tilde{s}}(\bar{x},\bar{y}=0)$ field (bottom left), and the energy density $\bar{\mathcal{E}}_{\rm int}(\bar{x},\bar{y}=0)$ (bottom middle) of two  superconducting string configurations with $\bar{\Tilde{s}}(r=0)=0.4$ (current density $J_{tot}=0.0155$) are shown for $\bar{d}=2$ with $\bar{\lambda}_\phi = 8,~\bar{\lambda}_\sigma = 80,~\bar{\beta} = 24$ and $\bar{\eta}_{\Tilde{\phi}}=0.55$.
}\label{fig:field configurations of two superconducting strings with current}
\end{figure}
\begin{figure}[t]
\centering
\includegraphics[width=15cm]{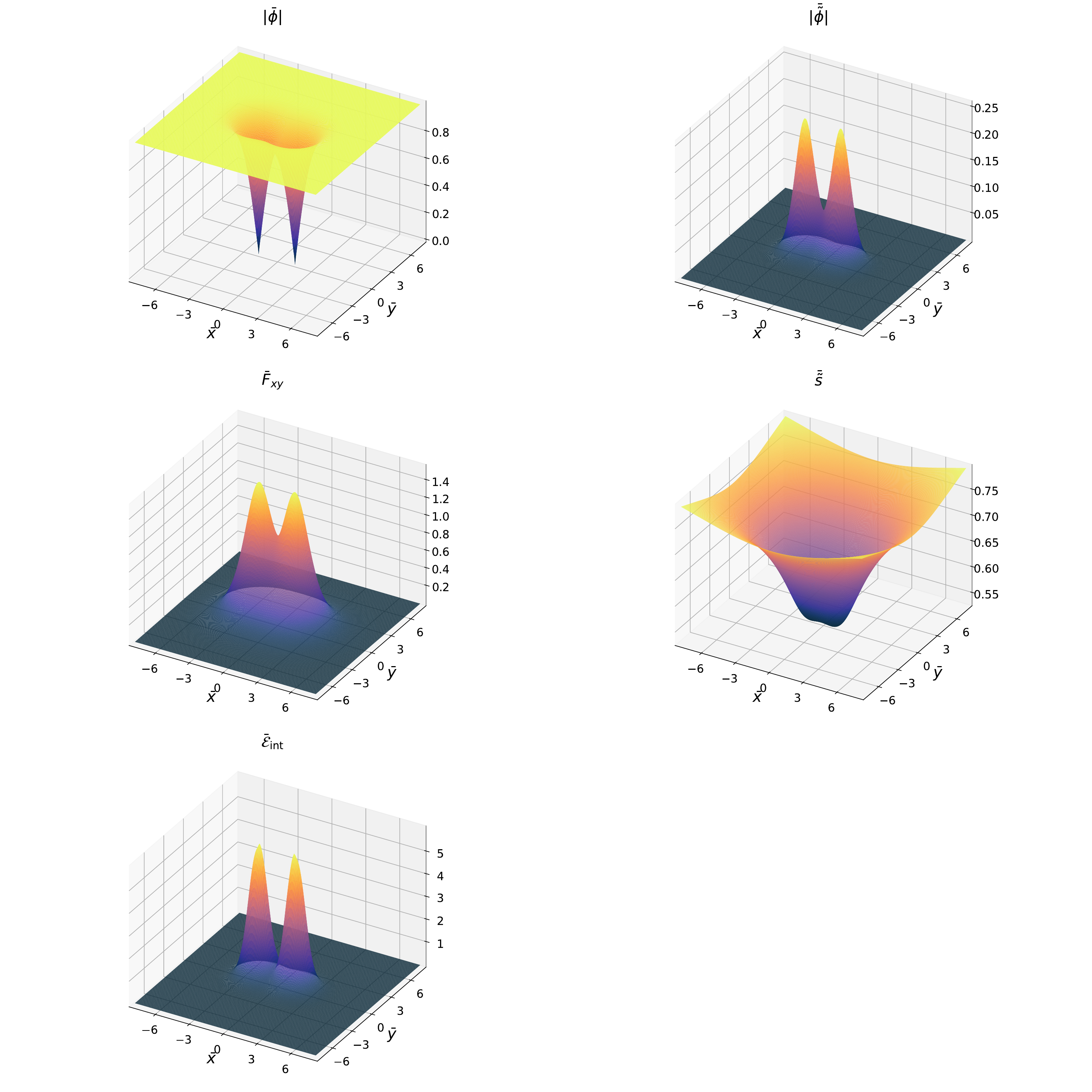}
\caption{Surface plots of the scalar field magnitude $|\bar{\phi}(\bar{x},\bar{y})|$ (top left), the $|\bar{\Tilde{\phi}}(\bar{x},\bar{y})|$ field (top right), the magnetic flux $\bar{F}_{xy}(\bar{x},\bar{y})$ (middle left), the $\bar{\Tilde{s}}(\bar{x},\bar{y})$ field (middle right), and the energy density $\bar{\mathcal{E}}_{\rm int}(\bar{x},\bar{y})$ (bottom left) of two superconducting string configurations with $\bar{\Tilde{s}}(r=0)=0.4$ (current density $J_{tot}=0.0155$)  are shown for $\bar{d}=2$ with $\bar{\lambda}_\phi = 8,~\bar{\lambda}_{\Tilde{\phi}} = 80,~\bar{\beta} = 24$ and $\bar{\eta}_{\Tilde{\phi}}=0.55$.
}\label{fig:field configurations of two superconducting strings with current 2D}
\end{figure}

Figure~\ref{fig:field configurations of two superconducting strings with current} shows the field configurations and energy density of two static superconducting strings on $\bar{y}=0$ slice with fixed separation, $\bar{d}=2$, for $\bar{\lambda}_\phi=8,~\bar{\lambda}_{\Tilde{\phi}}=80,~\bar{\beta} =24$,~$\bar{\eta}_{\Tilde{\phi}}=0.55$,~$\bar{\Tilde{s}}(0)=0.4$, while Fig.~\ref{fig:field configurations of two superconducting strings with current 2D} is a surface plot for the same parameter set.
We numerically confirm that the whole field configuration of $\widetilde{U}(1)$ gauge field can be approximated by a superposition of each string when two strings are far enough from each other.

\begin{figure}[t]
\centering
\includegraphics[width=15cm]{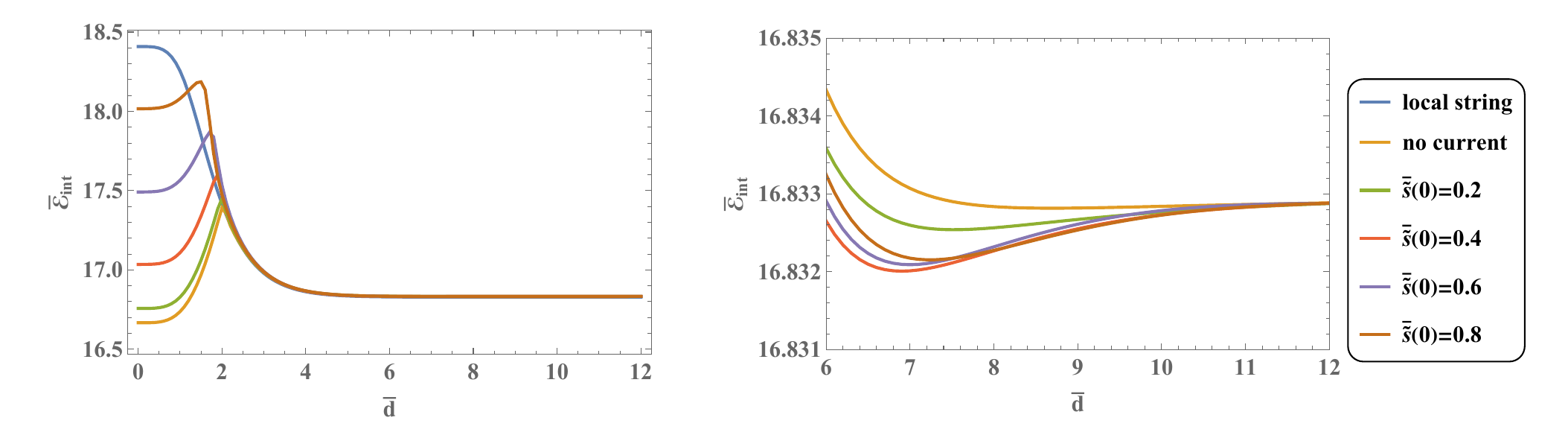}
\caption{
The dependence of the interaction energy of bosonic superconducting strings, $\bar{\mathcal{E}}_{\rm int}$, on the fixed distance, $\bar{d}$, are shown with $\bar{\lambda}_{\phi} = 8,~\bar{\lambda}_{\Tilde{\phi}} = 80,~\bar{\beta} = 24,~\bar{\eta}_{\Tilde{\phi}}=0.55$ for various initial values of $\bar{\Tilde{s}}$ (corresponding current density $J_{\rm tot}$ are shown in Table \ref{table:charge fitting}). The vacuum energy is unified with local string case for demonstration purpose.
}\label{fig:superconducting_interaction energies_scan2}
\end{figure}

Figure~\ref{fig:superconducting_interaction energies_scan2} shows the dependence of $\bar{\mathcal{E}}_{int}$ of superconducting strings on a separation distance $\bar{d}$ with various currents, including zero current.
For clarity, all vacuum energies are matched at $\bar{d}=12$.
The logarithmic growth of the $\widetilde{U}(1)$ gauge field can be seen in the long-distance limit, but it is very hard to see because of the small charge $k_A$ due to current quenching effects as explained in Sec.~\ref{sec:bosonic superconducting}.
In fact, the value of $k_A$ is tiny on the order of $10^{-2}\sim 10^{-3}$, while the other charges are order unity as shown in Table~\ref{table:charge fitting}.

As we have shown for the case without currents, the gradient energy of $\bar{\tilde{\phi}}$ leads to an additional attraction at short range.
If the amplitude of the $\widetilde{U}(1)$ gauge field, $\bar{\Tilde{s}}$, on the string axis is not zero, the $\bar{\tilde{\phi}}_r$ condensation becomes smaller, and consequently, the attraction becomes weaker than in the absence of current, as discussed in the previous section.
As the initial value of $\bar{\Tilde{s}}$ increases, the strength of the attraction becomes even weaker.
We can see this clearly from Fig.~\ref{fig:superconducting_interaction energies_scan2}.
This effect can be explained by the back-reaction on $\tilde{\phi}_r$, which is the same mechanism as that of current quenching.
Thus, even if an additional attraction due to the $\widetilde{U}(1)$ gauge field is introduced, the total attraction at short range is still suppressed by the back-reaction effect.
This implies that the attraction in the $\widetilde{U}(1)$ sector is maximized when the current is zero at a separation distance comparable to the string thickness.

\section{Conclusions}\label{sec:conclusion}

In this paper, we investigated the interaction energy of two static, straight, various types of cosmic strings using the point source formalism proposed in Ref.~\cite{Speight:1996px} and numerical calculations using the gradient flow method.
In the analysis based on the point source formalism, we analytically derived the asymptotic configurations of the cosmic strings and showed that the asymptotic field configurations can be realized by the linearized field theory introducing appropriate external sources.
The interaction energy can then be calculated using the standard Green's function method under the assumption that the external source can be approximated by a superposition of each string.
We confirmed that our results are consistent with those of previous studies on local~\cite{Speight:1996px} and global strings~\cite{Perivolaropoulos:1991du}.
In the case of bosonic superconducting strings, the Higgs field in the fundamental representation under the new gauge group $\widetilde{U}(1)$ is an additional source of attraction.
In addition to this, if the static magnetic current is trapped inside the string, it becomes a source of long-range attraction (repulsion) if the directions of the currents in the two strings are parallel (antiparallel).
However, due to the current quenching effect, there is an upper bound on the current, which leads to the upper bound on the charge of the string.
As a result, the $\widetilde{U}(1)$ gauge field cannot be the most significant source of interaction unless the strings are very far apart.
This is the essential reason why the apparent discontinuity (kink behavior) of the first derivative of $\bar{\mathcal{E}}_{\rm int}$ at the transition point is triggered by the coalescence of $\bar{\tilde{\phi}}_r$ rather than that of the $\widetilde{U}(1)$ gauge field, $\bar{\tilde{s}}$. 

We numerically evaluated the interaction energies of several types of cosmic strings at arbitrary separation distance using the relaxation method (gradient flow method), and compared the results with those obtained using the point source formalism.
In general, analyses relying on the point source formalism are in good agreement with the full numerical results of the gradient flow method only for separation distances longer than the inverse of the lightest mass in the underlying theory evaluated at infinity.
The results for the simple local string are in good agreement with the results of the previous study~\cite{Eto:2022hyt}, and with those by the variational approach~\cite{Jacobs:1978ch}.
Long-range interactions due to NG-bosons are also found to exist in the case of global strings.
For both local and global strings, the phase structure is simple in the sense that the interaction energy is a monotonic function of the separation distance.
In the case of superconducting strings, an additional short-range attraction is induced by the Higgs field in the fundamental representation of $\widetilde{U}(1)$.
This leads to the emergence of non-trivial phase structures of interaction energy (see {\it e.g.} Fig.~\ref{fig:superconducting_interaction energies_scan1}), that cannot be captured by the point source formalism.
Furthermore, when the effects of magnetic currents are included, a logarithmic dependence of the interaction energy on the separation distance on at large distance is found, which is consistent with the predictions of the point source formalism.
However, the total strength of the attraction is weakened by the current quenching effect at a separation distance comparable to the string thickness.
Therefore, we conclude that the attraction between two static bosonic superconducting strings is maximized when the total current associated with $\widetilde{U}(1)$ is zero.

Our new results on the interaction energy of bosonic superconducting strings indicate that there is a viable parameter region leading to an attraction force, at least when two strings are static and straight.
Thus, the necessary condition for the bound state of the two strings is satisfied.
Therefore, it would be very interesting to investigate the dynamical collision of the two-string system and to discuss the possibility of bound state formation of the two bosonic superconducting strings.
Such analysis could include the effects of non-zero relative angles and velocities on the bound state formation of the strings. We leave it as a future study.

\section*{Acknowledgement}
We would like to appreciate Yoshihiko Abe, Jinno Ryusuke, and Yu Hamada for valuable discussions.
We are grateful for fruitful discussions with Takashi Hiramatsu and Daisuke Yamauchi when this work is initiated.
KF is supported by JSPS Grant-in-Aid for Research Fellows Grant No.22J00345.
S.L. is supported by JSPS Grant-in-Aid for Research Fellows Grant No.23KJ0936.
M.Y. is supported by IBS under the project code, IBS-R018-D3, and by JSPS Grant-in-Aid for Scientific Research Number JP21H01080.
\appendix

\section{Normalization and Numerical solutions}\label{appendix:numerical calculation}

This appendix describes the normalization used in the numerical calculations and the numerical recipes for obtaining various types of cosmic string configurations.

\subsection{Normalization of local and bosonic superconducting strings} \label{appendix:local normalization}
We consider the following action:
\begin{equation}
\begin{split}
    &S = \int {\rm d}^4x\left[-\frac{1}{4}F^{\mu\nu}F_{\mu\nu} + |D_\mu \phi|^2 -\frac{1}{4}\widetilde{F}^{\mu\nu}\widetilde{F}_{\mu\nu} +|\widetilde{D}_\mu \Tilde{\phi}|^2 - V(\phi,\Tilde{\phi})\right],\\
    &V(\phi,\Tilde{\phi}) = \frac{\lambda_\phi}{4} \left( |\phi|^2-\eta_\phi^2 \right)^2 + \frac{\lambda_{\Tilde{\phi}}}{4}\left(|\Tilde{\phi}|^2-\eta_{\Tilde{\phi}}^2\right)^2+\beta |\phi|^2 |\Tilde{\phi}|^2 .\label{eq:action superconducting}
\end{split}
\end{equation}
For convenience, we introduce the following dimensionless length $\bar{x}$ and various dimensionless fields $\bar{\phi},~\bar{A}_\mu,~\bar{\Tilde{\phi}}$ and $\bar{\Tilde{A}}_\mu$ as  
\begin{align}
    \bar{x}\equiv e\eta_\phi x,~\bar{\phi} \equiv\dfrac{\phi}{\eta_\phi},~\bar{A}_\mu \equiv\dfrac{A_\mu}{\eta_\phi},~\bar{\Tilde{\phi}}\equiv \dfrac{\Tilde{\phi}}{\eta_\phi},~\bar{\Tilde{A}}_\mu \equiv \dfrac{\Tilde{A}_\mu}{\eta_\phi}\,. \label{eq:normalization}
\end{align}
Then the action Eq.~\eqref{eq:action superconducting} becomes 
\begin{align}
     &S = \dfrac{1}{e^2}\bar{S}, \nonumber \\
     &\bar{S}\equiv\int {\rm d}^4\bar{x}\left[-\frac{1}{4}\bar{F}^{\mu\nu}\bar{F}_{\mu\nu} + |\bar{D}_\mu \bar{\phi}|^2 -\frac{1}{4}\bar{\Tilde{F}}^{\mu\nu}\bar{\Tilde{F}}_{\mu\nu} +|\bar{\Tilde{D}}_\mu \bar{\Tilde{\phi}}|^2 - \overline{V}\left(\bar{\phi},\,\bar{\Tilde{\phi}}\right)\right],\\
     &\overline{V}\left(\bar{\phi},\,\bar{\Tilde{\phi}}\right) \equiv \frac{\bar{\lambda}_\phi}{4} \left( |\bar{\phi}|^2-1  \right)^2 + \frac{\bar{\lambda}_{\Tilde{\phi}}}{4}\left(|\bar{\Tilde{\phi}}|^2-\bar{\eta}_{\Tilde{\phi}}^2\right)^2+\bar{\beta} |\bar{\phi}|^2 |\bar{\Tilde{\phi}}|^2 \nonumber \,.
\end{align}
where $\bar{F}_{\mu\nu}\equiv \bar{\partial}_\mu \bar{A}_\nu-\bar{\partial}_\nu \bar{A}_\mu,~\bar{\Tilde{F}}_{\mu\nu}\equiv \bar{\partial}_\mu \bar{\Tilde{A}}_\nu-\bar{\partial}_\nu \bar{\Tilde{A}}_\mu,~\bar{D}_\mu \equiv \bar{\partial}_\mu-i\bar{A}_\mu$ and $\bar{\Tilde{D}}_\mu = \bar{\partial}_\mu-i\bar{g} \bar{A}_\mu$ with $\bar{\partial}_\mu\equiv \partial/\partial \bar{x}^\mu$.
The physical parameters in the above expressions are defined by 
\begin{align}
     \bar{\eta}_{\Tilde{\phi}}\equiv \dfrac{\eta_{\Tilde{\phi}}}{\eta_\phi},~\bar{\lambda}_\phi \equiv \frac{\lambda_\phi}{e^2},~\bar{\lambda}_{\Tilde{\phi}}\equiv \dfrac{\bar{\lambda}_{\Tilde{\phi}}}{e^2},~\bar{\beta} \equiv \frac{\beta}{e^2},~\bar{g}\equiv \frac{g}{e}\,.\label{eq:coupling normalization}
\end{align}
Assuming a static configuration, the two-dimensional energy (tension) of the string can be extracted as follows,
\begin{equation}
\begin{split}
	&\mathcal{E}_{\rm int}= \frac{1}{e} \bar{\mathcal{E}}_{\rm int},~\bar{\mathcal{E}}_{\rm int} \equiv \int {\rm d}^2\bar{x}\,\epsilon_{\rm int}(\bar{x},\bar{y}),\\
	&\epsilon_{\rm int}(\bar{x},{\bar{y}})\equiv \left[\frac{1}{4}\bar{F}^{ij}\bar{F}_{ij} + |\bar{D}_i \bar{\phi}|^2 +\frac{1}{4}\bar{\Tilde{F}}^{ij}\bar{\Tilde{F}}_{ij} +|\bar{\Tilde{D}}_i \bar{\Tilde{\phi}}|^2 + \overline{V}\left(\bar{\phi},\,\bar{\Tilde{\phi}}\right)\right].\label{eq:normalized two dimensional Eint}
\end{split}
\end{equation}
In this normalization, $\bar{\eta}_{\Tilde{\phi}},~\bar{\lambda}_\phi,~\bar{\lambda}_{\Tilde{\phi}},~\bar{\beta}$ and $\bar{g}$ are treated as free parameters.
Note that this normalization cannot be applied to global strings because $\bar{x}$ becomes ill-defined for $e=0$. Therefore, we will use a different normalization for global strings.

\subsection{Normalization of global strings} \label{appendix:global normalization}

In the case of global strings, the normalization defined by Eqs.~\eqref{eq:normalization} and \eqref{eq:coupling normalization} cannot take the limit $e\to 0$, so the normalization of the complex scalar field and length scale must be changed.
We then use the following normalization in the numerical computation of the global strings.
\begin{align}
\bar{\phi}\equiv \dfrac{\phi}{\eta_\phi},~\bar{x}\equiv \eta_\phi x\,.
\end{align}
With this normalization, the tension of the global string becomes
\begin{align}
&\bar{\mathcal{E}}_{\rm int} = \int {\rm d}^2\bar{x}\left( |\bar{\partial}_\mu \bar{\phi}|^2-V(\bar{\phi})\right) ,\\
    &V(\bar{\phi}) = \frac{\lambda_\phi}{4} \left( |\bar{\phi}|^2- 1\right)^2 .
\end{align}

\subsection{Numerical solutions}

Here, we describe a numerical recipe for obtaining the string configurations.
The cosmic string solution is obtained by solving the coupled second-order differential equations defined by Eqs.~\eqref{eq:bosonic eom}.
The boundary conditions for these differential equations at infinity are given by Eq.~\eqref{eq:superconducting string boundary condition}.
As in the case of the conventional ANO strings, the regularity of $\phi_r(r)$ and $A_\theta(r)$ at the origin requires the conditions given by Eq.~\eqref{eq:ANO boundary conditions at the origin}.
Also, $\partial_r {\Tilde{\phi}}_r (r)$ and $\partial_r \Tilde{s}(r)$ must vanish at the origin by these regularities.
In addition, in order to solve the equation of $\tilde{s}(r)$, we take another boundary condition $\tilde{s}(r=0)$, which is considered crucial for the formation of bosonic superconducting strings with current.
In this paper, we treat $\Tilde{s}(r=0)$ as a free parameter and simply read off the total current flowing inside the current using Eq.~\eqref{eq:superconducting current}.

Since the boundary conditions for $\phi_r(r),~A_\theta (r)$ and ${\Tilde{\phi}}_r (r)$ are given both at infinity and at the origin, it is a boundary value problem to solve these fields.
On the other hand, the two boundary conditions for $\Tilde{s}$ are specified both at the origin, which is a standard initial value problem. 
Hence we apply the gradient flow method to find solutions of $\phi_r(r),~A_\theta (r)$, and ${\Tilde{\phi}}_r (r)$, while the standard numerical integration method is applied for $\Tilde{s}$.
For other numerical approaches, including variational analysis and successive over-relaxation method, we refer to Refs.~\cite{Adler:1983zh,Hill:1987qx,Amsterdamski:1988zp,Davis:1988jp}.

In the gradient flow method, the $\phi_r (r),~A_\theta(r)$ and ${\Tilde{\phi}}_r (r)$ fields are promoted to fictitious time-dependent fields as $\phi_r(t,r),~A_\theta(t,r)$ and ${\Tilde{\phi}}_r(t,r)$.
Using the normalization defined in Sec.~\ref{sec:numerical study}, we consider the following heat equations.
\begin{align}
    &\frac{\partial^2 \bar{\phi}_r}{\partial \bar{r}^2} + \frac{1}{\bar{r}}\frac{\partial \bar{\phi}_r}{\partial \bar{r}} - \frac{1}{\bar{r}^2}\bar{\phi}_r\left( \bar{A}_\theta - n \right)^2 - \frac{1}{2}\bar{\lambda}_\phi \bar{\phi}_r (\bar{\phi}_r^2 - 1) - \bar{\beta}|{\bar{\Tilde{\phi}}}|^2\bar{\phi}_r = \dfrac{\partial \bar{\phi}_r}{\partial t}, \nonumber\\
    &\frac{\partial^2 \bar{A}_\theta}{\partial \bar{r}^2} - \frac{1}{\bar{r}}\frac{\partial \bar{A}_\theta}{\partial \bar{r}} - 2 \phi_r^2 \left( \bar{A}_\theta - n \right) = \dfrac{\partial \bar{A}_\theta }{\partial t},\label{eq:heat equation}\\
    &\frac{\partial^2 \bar{\Tilde{\phi}}_r}{\partial \bar{r}^2} + \frac{1}{\bar{r}}\frac{\partial \bar{\Tilde{\phi}}_r}{\partial \bar{r}} - \bar{\Tilde{s}}(\bar{r})^2{\Tilde{\phi}}_r - \frac{1}{2} \bar{\lambda}_{\Tilde{\phi}}{\bar{\Tilde{\phi}}}_r({\bar{\Tilde{\phi}}}_r^2 - \bar{\eta}_{\Tilde{\phi}}^2) - \bar{\beta}|\bar{\phi}|^2{\bar{\Tilde{\phi}}}_r = \dfrac{\partial {\bar{\Tilde{\phi}}}_r}{\partial t}.\nonumber
\end{align}
Here, the signs of the time derivatives of the various fields are determined such that the energy of the system decreases with time evolution.
First, we set $\bar{\Tilde{s}}(\bar{r})=0$ and find the field configurations for $\bar{\phi},~\bar{A}_\theta$ and $\Tilde{\phi}$.
We initially prepare functions for the fields  $\bar{\phi}_r (t=0,\bar{r}),~\bar{A}_\theta(t=0,\bar{r})$ and $\bar{{\Tilde{\phi}}}_r(t=0,\bar{r})$ that satisfy boundary conditions at origin and infinity, then evolve these functions numerically with the fictitious time.
Specifically, for $\bar{\phi}_r$ and $\bar{A}_\theta$, we prepare hyperbolic tangent configuration, and Gaussian configurations for $\bar{\Tilde{\phi}}_r$.
When the time evolution converges to $\partial_t \bar{\phi}_{\bar{r}}=\partial_t \bar{A}_\theta=\partial_t{\bar{\Tilde{\phi}}}_r=0$, a solution to the original differential equations is obtained.
The time evolution converges quickly, consistent with the results of Ref.~\cite{Battye:2021kbd}.

Now let us include the effect of the gauge field, $\Tilde{s}(r)$.
In addition to the differential equations Eq.~\eqref{eq:heat equation}, we need to include the following differential equation,
\begin{align}
\frac{\partial^2 \bar{\Tilde{s}}}{\partial \bar{r}^2} + \frac{1}{\bar{r}}\frac{\partial \bar{\Tilde{s}}}{\partial \bar{r}} - 2g^2\bar{\Tilde{s}}(\bar{r}){\bar{\Tilde{\phi}}}_r^2 = 0\,.\label{eq:heat equation current}
\end{align}
The differential equations Eqs.~\eqref{eq:heat equation} and \eqref{eq:heat equation current} are solved iteratively.
To be precise, the following procedure is used for our numerical computation.
\begin{itemize}
	\item (i) We first obtain the configuration of $\bar{\phi}_r,~\bar{A}_\theta$ and $\bar{\Tilde{\phi}}_r$ by solving Eq.~\eqref{eq:heat equation} without the gauge field, $\bar{\Tilde{s}}(\bar{r})=0$, using the gradient flow method.
	\item (ii) Then, we solve Eq.~\eqref{eq:heat equation current} with the $\bar{\Tilde{\phi}}_r$ configuration evaluated in the first step (i) (or in the third step (iii)) using the standard initial value method. The initial value of $\bar{\tilde{s}}(\bar{r}=0)$ is set by hand.
	\item (iii) We use the current $\bar{\Tilde{s}}(\bar{r})=0$ evaluated in the second step and solve Eq.~\eqref{eq:heat equation} again to update the configuration of $\bar{\phi}_r,~\bar{A}_\theta$ and $\bar{\Tilde{\phi}}_r$. After this, we return to the second step (ii).
\end{itemize}
If the above iteration converges well, the whole configuration of the bosonic superconducting string, including the gauge field, can be obtained numerically.
We have confirmed that the iteration converges well, at least for the benchmark points shown in Table.~\ref{table:charge fitting}.

\begin{table}[!t]
    \centering
    \begin{tabular}{|c|c|c|c|c|c|c|c|}
    \hline
      &\makecell[c]{$\left(\bar{\lambda}_\phi, \bar{g}, \bar{\lambda}_{\Tilde{\phi}}, \bar{\beta}, \bar{\eta}_{\Tilde{\phi}}\right)$} & \makecell[c]{$\bar{\Tilde{s}}(\bar{r}=0)$} &  $\widetilde{J}_{\rm tot}$  & $k_\phi$ & $k_e$ & $k_{\Tilde{\phi}}$ & $k_A$\\ 
     \hline
     & & 0.1 & 0.0185 & 2.2332 & 2.2516 &0.8026 & 0.0059\\
     && 0.2 & 0.0277& 2.1675 & 2.2375 &0.6812 &0.0088\\
     $A$&{(2.8, $\sqrt{2}$, 20, 6, 0.61)}& 0.3 & 0.0209& 2.0731 &2.2165 &0.4849 &0.0067\\
     &&0.35 &0.0104& 2.0195 &2.2043 &0.3230 &0.0033\\
     &&0.378 & 0.0017 & 1.9862 &2.1966 &0.1293 &0.0006\\
     \hline
      && 0.7& 0.0885 & 3.7869 & 1.9155 &1.8362 & 0.0282\\
     && 1.0& 0.1014 & 3.7036 & 1.9061 &1.1889 & 0.0323\\
     $B$&{(6, $\sqrt{6}$, 216, 39.6, 0.4)}& 1.3 & 0.0940 & 3.6119 & 1.8956 &0.7691 &0.0299\\
     && 1.6 & 0.0670 & 3.5175 & 1.8855 &0.4574 &0.0213\\
     &&1.913&0.0094 & 3.4151 & 1.8729 &0.1821 &0.0030\\
    \hline
     && 0.2& 0.0091 & 5.4473 & 1.8078 &1.8810 & 0.0029\\
     && 0.4& 0.0155 & 5.3950 & 1.8048 &1,7241 & 0.0049\\
     $C$&{(8, 1, 80, 24, 0.55)}& 0.6 & 0.0166 & 5.3105 & 1.7991 &1.4397 &0.0053\\
     && 0.8 & 0.0105 & 5.2054 & 1,7917 &0.9842 &0.0033\\
     &&0.920&0.0018 & 5.1285 & 1.7862 &0.3826 &0.0006\\
     \hline 
    \end{tabular}
    \caption{
   Numerical constants $k_\phi,~k_e,~k_{{\Tilde{\phi}}},~k_A$, and $\widetilde{J}_{\rm tot}$ are computed for three benchmark points $A,~B$ and $C$. Definitions of these constants can be found by Eqs.~\eqref{eq:asymptotic sigma} \eqref{eq:asymptotic U} \eqref{eq:asymptotic tildephi} and \eqref{eq:asymptotic gauge field}.}\label{table:charge fitting}
\end{table}

The effective origin and infinity are set to $\bar{r}=10^{-10}$ and $\bar{r}=40$, respectively, and the finite size effect is found to be negligible.
We take three benchmark points $A,~B$ and $C$ shown in Table \ref{table:charge fitting}.
The numerical constants $k_\phi,~k_e,~k_{\Tilde{\phi}}$ and $k_A$ are determined such that the asymptotic solutions Eqs.~\eqref{eq:asymptotic sigma} \eqref{eq:asymptotic U} \eqref{eq:asymptotic tildephi} and \eqref{eq:asymptotic gauge field} are matched with the numerical results at infinity.
From Table.~\ref{table:charge fitting}, it is easy to see that the total current calculated in Eq. \eqref{eq:superconducting current} decreases as $\Tilde{s}(r=0)$ is increased above the threshold value.

\begin{figure}[t]
\centering\includegraphics[width=7.5cm]{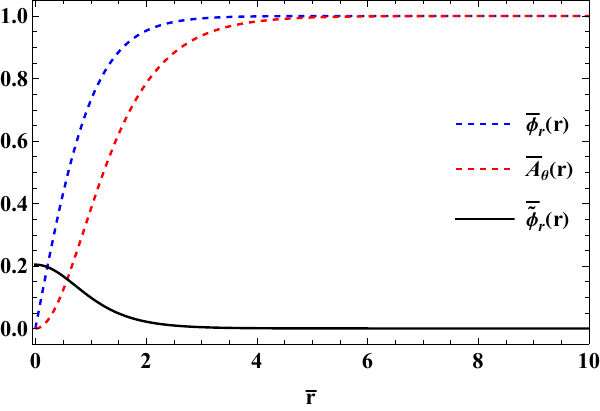}
\includegraphics[width=7.5cm]{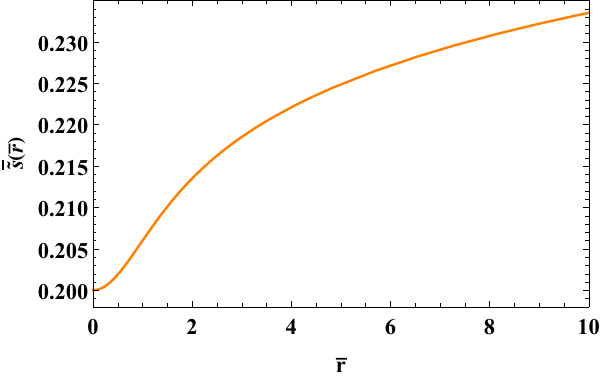}
\caption{
$\bar{\phi}_r(\bar{r}),~\bar{A}_\theta(\bar{r}),~\bar{\Tilde{\phi}}_r(\bar{r})$ configurations of a bosonic superconducting string are shown (left panel) for the benchmark point $A$ with $\bar{\Tilde{s}}(\bar{r}=0)=0.2$. Also shown is the configuration of $\bar{\Tilde{s}}(\bar{r})$ (right panel).}\label{fig:superconducting solution}
\end{figure}
\begin{figure}[t]
\centering\includegraphics[width=7.5cm]{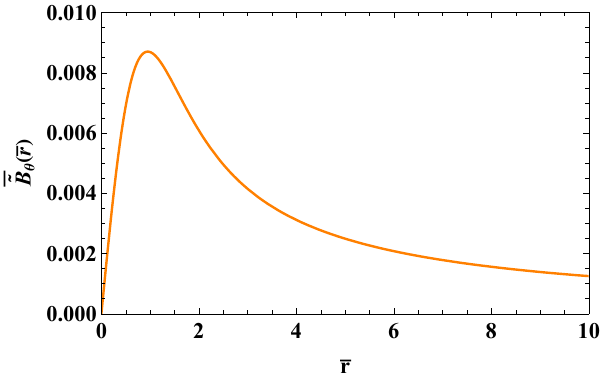}
\caption{
The magnetic field $\bar{\Tilde{B}}_\theta=\partial_{\bar{r}} \bar{\Tilde{A}}_z(\bar{r})$ for the benchmark point $A$ with the initial condition $\bar{\Tilde{s}}(\bar{r}=0)=0.2$ is shown.
}\label{fig:magnetic field}
\end{figure}
The configurations of $\bar{\phi}_r,~\bar{A}_\theta,~\bar{\Tilde{\phi}}$ and $\bar{\Tilde{s}}$ are shown in Fig.~\ref{fig:superconducting solution} for the benchmark point $A$ with initial condition $\bar{\Tilde{s}}=0.2$ (leading to the total current $\widetilde{J}_{\rm tot}\simeq 0.0277$).
As expected, the configurations of $\bar{\phi}_r (\bar{r}),~\bar{A}_\theta (\bar{r})$ and $\bar{\Tilde{\phi}}_r$ are exponentially suppressed, while $\bar{\Tilde{s}}(\bar{r})$ increases logarithmically at long distance.
The corresponding magnetic field defined as $\bar{\Tilde{B}}_\theta=\partial_{\bar{r}}\bar{\Tilde{A}}_z(\bar{r})$ is depicted in Fig.~\ref{fig:magnetic field}.
It is clear from the figure that the magnetic field is proportional to $1/\bar{r}$ outside the string but decays towards the center of string, where it decays to zero.
The penetration depth of the magnetic field is approximately given by the mass of the gauge field acquired by the Higgs mechanism inside the string.
This effect is known as the {\it Meissner effect}.

\section{Numerical calculation of the interaction energy of two cosmic strings}\label{appendix:interaction energy}

This appendix details the numerical setup for the interaction energy of two straight cosmic strings.
In the system of two straight cosmic strings, the calculations are performed in Cartesian coordinates rather than cylindrical coordinates, since the cylindrical symmetry is explicitly broken unless the axes of the two strings coincide.
To find the configuration, we apply the gradient flow method as in the case of a single straight string.
In particular, we parameterize the fields $\bar{\phi},~\bar{A}_\mu,~\bar{\Tilde{\phi}}$ and $\bar{\Tilde{s}}$ as 
\begin{equation}
\begin{split}	
	&\bar{\phi}(t,x,y) = \bar{\phi}_x(t,x,y) +i\bar{\phi}_y (t,x,y)\, , ~
	\bar{A}_\mu (t,x,y) = (0,\bar{A}_x (t,x,y),\bar{A}_y (t,x,y),0)\,,\\
	&\bar{\Tilde{\phi}}(t,x,y) = \bar{\Tilde{\phi}}_r (t,x,y)\,,~
	\bar{\Tilde{A}}_z (t,x,y) = \frac{1}{\bar{g}}\bar{\Tilde{s}}(t,x,y)\,.
\end{split}
\end{equation}
Since we work in the Coulomb gauge, an additional term $\mathcal{L}_{\rm Coulomb}=(\bar{\partial}_x \bar{A}_x+\bar{\partial}_y \bar{A}_y)^2/2$ is added to the Lagrangian density defined by Eq~\eqref{eq:superconducting lagrangian density}.
Using this parameterization and the normalization defined in Sec.~\ref{sec:numerical study}, we numerically solve the following equations,
\begin{equation}
\begin{split}	
	&\dfrac{\partial \bar{\phi}_x}{\partial t}=\bar{\Delta} \bar{\phi}_x   - |\bar{A}|^2\bar{\phi}_x  +2\left(\bar{A}_x\dfrac{\partial \bar{\phi}_x}{\partial \bar{x}}+\bar{A}_y\dfrac{\partial \bar{\phi}_y}{\partial \bar{y}}\right) -\dfrac{\bar{\lambda}_\phi}{2}\bar{\phi}_x\left(|\bar{\phi}|^2- 1\right)-\bar{\beta} \,\bar{\Tilde{\phi}}_r^2\bar{\phi}_x\, ,\\
	&\dfrac{\partial \bar{\phi}_y}{\partial t}=\bar{\Delta} \bar{\phi}_y    - |\bar{A}|^2 \bar{\phi}_y  -2\left(\bar{A}_x\dfrac{\partial \bar{\phi}_x}{\partial \bar{x}}+\bar{A}_y\dfrac{\partial \bar{\phi}_y}{\partial \bar{y}}\right) -\dfrac{\bar{\lambda}_\phi}{2}\bar{\phi}_y\left(|\bar{\phi}|^2 - 1\right)  -\bar{\beta}\, \bar{\Tilde{\phi}}_r^2\bar{\phi}_y\,, \\
	 &\dfrac{\partial \bar{A}_x}{\partial t} = \bar{\Delta} \bar{A}_x - 2|\bar{\phi}|^2\bar{A}_x+2\left(\bar{\phi}_x \dfrac{\partial \bar{\phi}_y}{\partial \bar{x}}-\bar{\phi}_y\dfrac{\partial \bar{\phi}_x}{\partial \bar{x}}\right)\,,\\
	 &\dfrac{\partial \bar{A}_y}{\partial t} = \bar{\Delta} \bar{A}_y - 2|\bar{\phi}|^2\bar{A}_y+2\left(\bar{\phi}_x \dfrac{\partial \bar{\phi}_y}{\partial \bar{y}}-\bar{\phi}_y\dfrac{\partial \bar{\phi}_x}{\partial \bar{y}}\right) \,,\\
	&\dfrac{\partial \bar{\Tilde{\phi}}_r}{\partial t}=\bar{\Delta} \bar{\Tilde{\phi}}  - \bar{\Tilde{s}}^2\bar{\Tilde{\phi}}_r - \frac{\bar{\lambda}_{\Tilde{\phi}}}{2}{\bar{\Tilde{\phi}}}_r(\bar{\Tilde{\phi}}_r^2 - \bar{\eta}_{\Tilde{\phi}}^2) - \bar{\beta}|\bar{\phi}|^2{\bar{\Tilde{\phi}}}_r\, , \\
	&\dfrac{\partial \bar{\Tilde{s}}}{\partial t}=\bar{\Delta} \bar{\Tilde{s}} - 2\bar{g}^2\,\bar{\Tilde{s}}\,\bar{{\Tilde{\phi}}}_r^2\,. \label{eq:2D time evolution}
\end{split}     
\end{equation}
In these expressions, $\bar{\Delta} \equiv \partial_{\bar{x}}^2+\partial_{\bar{y}}^2$ and $|\bar{A}|^2\equiv \bar{A}_x^2+\bar{A}_y^2$ are the standard two-dimensional Laplace operator and the two-dimensional norm for $U(1)$ gauge fields, respectively.

At initial time $\bar{t}=0$, all field configurations are taken to be
\begin{equation}
\begin{split}
	&\bar{\phi}(\bar{t}=0,\bar{x},\bar{y}) = F(\bar{x},\bar{y})e^{i\theta_1+i\theta_2},\\
	&\bar{A}_x (\bar{t}=0,\bar{x},\bar{y}) = F(\bar{x},\bar{y}) \left(a_{\bar{x}_+}(\bar{x},\bar{y}) + a_{\bar{x}_-}(\bar{x},\bar{y})\right),\\ 
	&\bar{A}_y (\bar{t}=0,\bar{x},\bar{y}) =F(\bar{x},\bar{y})\left(a_{\bar{y}_+}(\bar{x},\bar{y}) + a_{\bar{y}_-}(\bar{x},\bar{y})\right), \\
	&F(\bar{x},\bar{y})\equiv \tanh\left(\sqrt{(\bar{x}-\bar{d})^2+\bar{y}^2}\right)\tanh\left(\sqrt{(\bar{x}+\bar{d})^2+\bar{y}^2}\right),\\
	& a_{\bar{x}_{\pm}}(\bar{x},\bar{y})\equiv
	\begin{cases}	
 \dfrac{\mp \bar{y}}{(\bar{x}\pm \bar{d})^2+\bar{y}^2}~&((\bar{x}\pm \bar{d})^2+\bar{y}^2>0)\\
~~~~~~~~0&(\bar{x}=\mp \bar{d},~\bar{y}=0)
 \end{cases},\\
 & a_{\bar{y}_{\pm}}(\bar{x},\bar{y})\equiv
	\begin{cases}	
 \dfrac{(\bar{x}\pm \bar{d})}{(\bar{x}\pm \bar{d})^2+\bar{y}^2}~&((\bar{x}\pm \bar{d})^2+\bar{y}^2>0)\\
~~~~~~~~0&(\bar{x}=\mp \bar{d},~\bar{y}=0)
 \end{cases},\\
	&\bar{\Tilde{\phi}}_r (\bar{t}=0,\bar{x},\bar{y}) = \dfrac{\bar{\eta}_{\Tilde{\phi}}}{2}\left(e^{-\sqrt{(\bar{x}-\bar{d})^2+\bar{y}^2}} +e^{-\sqrt{(\bar{x}+\bar{d})^2+\bar{y}^2} }\right).\\
	&\bar{\Tilde{s}}=\bar{\Tilde{s}}^{\rm inf}_1 + \bar{\Tilde{s}}^{\rm inf}_2
\end{split} 
\end{equation}
Here, $\theta_1$ and $\theta_2$ are the azimuthal angles measured from each string axe placed at $(\bar{x},\bar{y})=(-\bar{d}/2,0)$, and $(\bar{x},\bar{y})=(\bar{d}/2,0)$, respectively.
$\bar{\tilde{s}}_i^{\rm{inf}}$ is the solution of the $\widetilde{U}(1)$ gauge field in the single string case obtained in App.~\ref{appendix:numerical calculation}.
It should be noted that above initial conditions satisfy the desired boundary conditions, $\bar{\phi}(\bar{x},\bar{y})\to  e^{i\theta_1+i\theta_2},~|\bar{D}_\mu \bar{\phi}(\bar{x},\bar{y})|\to 0,~\bar{\Tilde{\phi}}_r(\bar{x},\bar{y})\to0$ at $\sqrt{\bar{x}^2+\bar{y}^2}\to \infty$, and $\bar{\phi}(\pm\bar{d}/2,0)=0$.
For local strings, $\bar{\beta}=\bar{\eta}_{\tilde{\phi}}=0$ and $\bar{\tilde{s}}_i^{\rm{inf}}=0$ are taken while $\bar{A}_x (\bar{t}=0,\bar{x},\bar{y}) = \bar{A}_y (\bar{t}=0,\bar{x},\bar{y}) = 0$ is additionally taken for global strings.

To solve the above equations numerically, we prepare a two-dimensional simulation box with a length of $L=30$ per side (the box size is increased to $L=50$ for superconducting strings with current.
The lattice spacing is $a=0.05$ in units of $e\eta_{\phi} = 1$ for local and bosonic superconducting strings and  $a=0.05$ in units of $\eta_{\phi}=1$ for global strings.
The (fictitious) time evolution of all fields is evaluated from $\bar{t}=0$ to $\bar{t}=15$ in time steps of $\delta \bar{t} =5\times 10^{-4}$ by the standard Euler method.

Since we would like to evaluate a static field configuration under a fixed separation, we need to fix the position of the string axis during the time evolution.
To do so, we also fix the phase factor of $\bar{\phi}$ at each time step according to Ref.~\cite{Jacobs:1978ch}.
Specifically, we first evaluate $\bar{\phi}(\bar{t}+\delta \bar{t},\bar{x},\bar{y})$ from $\bar{\phi}(\bar{t},\bar{x},\bar{y})$ by Eq.~\eqref{eq:2D time evolution}, and then, it is updated by replacing it with $\bar{\phi}(\bar{t}+\delta \bar{t},\bar{x},\bar{y})\to |\bar{\phi}(\bar{t}+\delta \bar{t},\bar{x},\bar{y})|e^{i\theta_1+i\theta_2}$.
In the string axis where the phase factors, $\theta_1$ and $\theta_2$, are ill-defined, the value of $\bar{\phi}$ is fixed at each time step as $\bar{\phi} = 0$.
We use the Julia programming language to solve the above partial differential equations with aforementioned setup in our numerical calculations.
After evolving the field configurations, the two-dimensional energy density of the two-string system defined by Eq.~\eqref{eq:normalized two dimensional Eint} is computed as 
\begin{equation}
\begin{split}
\bar{\mathcal{E}}_{\rm int} = \int \mathrm{d}x \mathrm{d}y \,& (\partial_{\bar{x}} \bar{\phi}_x + \bar{A}_x\bar{\phi}_y)^2 + (\partial_{\bar{x}}\bar{\phi}_y - \bar{A}_x\bar{\phi}_x)^2 + (\partial_{\bar{y}}\bar{\phi}_x + \bar{A}_y\bar{\phi}_y)^2 + (\partial_{\bar{y}}\bar{\phi}_y - \bar{A}_y\bar{\phi}_x)^2  \\
&+\frac{1}{2}(\partial_{\bar{x}}\bar{A}_y - \partial_{\bar{y}} \bar{A}_x)^2 + \frac{1}{2}(\bar{\partial}_{\bar{x}} \bar{A}_x+\bar{\partial}_{\bar{y}} \bar{A}_y)^2 + (\partial_{\bar{x}} \bar{\tilde{\phi}}_r)^2 + (\partial_{\bar{y}} \bar{\tilde{\phi}}_r)^2 + \bar{g}^2\bar{\tilde{s}}^2\bar{\tilde{\phi}}_r^2 +\\
&\frac{1}{2}(\partial_{\bar{x}}\bar{\tilde{s}})^2 + \frac{1}{2}(\partial_{\bar{y}}\bar{\tilde{s}})^2+\bar{V}(\bar{\phi},\bar{\tilde{\phi}})],\\
\bar{V}(\bar{\phi},\bar{\tilde{\phi}}) = &\frac{\bar{\lambda}_\phi}{4}(\bar{\phi}_x^2 + \bar{\phi}_y^2 - 1)^2 +\frac{\bar{\lambda}_{\Tilde{\phi}}}{4} (\bar{\Tilde{\phi}}_r^2 - \bar{\eta}_{\Tilde{\phi}}^2)^2 + \bar{\beta} (\bar{\phi}_x^2 + \bar{\phi}_y^2)\bar{\Tilde{\phi}}_r^2 - \frac{\bar{\lambda}_{\Tilde{\phi}}}{4} \bar{\eta}_{\Tilde{\phi}}^2.
\end{split}
\end{equation}

\newpage

\bibliography{axion_string}

\end{document}